\documentstyle[12pt]{article}
\def\beq{\begin{equation}}
\def\eeq{\end{equation}}
\begin{document}

\begin{titlepage}

%\vspace{1cm}\\

\centerline{\large \bf  Effective action for the Regge processes in gravity}

\date{}

\begin{center}
L.N. Lipatov {%$^{*}$
\\
St. Petersburg State University,\\
Petersburg Nuclear Physics Institute,\\
Gatchina, 188300, St.Petersburg, Russia,\\
Universit\"{a}t Hamburg,\\
II. Institut f\"{u}r Theoretische Physik,\\
Luruper Chaussee,149, D-22761 Hamburg}
\end{center}
\vskip 15.0pt
\vspace{-7cm}
\begin{flushright}
%$~~$\\
DESY-11-065
\end{flushright}
\vspace{8cm}

\centerline{\bf Abstract}

\noindent

It is shown, that the  effective action for the reggeized graviton interactions
can be formulated in terms of the reggeon fields $A^{++}$ and $A^{--}$ and the metric tensor
$g_{\mu \nu}$
in such a way, that it is local in the rapidity space and has the property of general
covariance. The corresponding effective currents $j^{-}$ and $j^{+}$
satisfy the Hamilton-Jacobi equation for a massless particle moving in the gravitational
field. These currents are calculated explicitly for the shock wave--like fields  and a
variation principle for them is formulated.  As an application, we reproduce the
 effective lagrangian for the
multi-regge processes in gravity together with the graviton Regge trajectory in the leading logarithmic
approximation with taking into account supersymmetric contributions.

\vskip 3cm

\hrule

\vskip 3cm

\noindent

%\noindent $^{*}$ {\it Supported by the RFBR grant} \vfill

\end{titlepage}

\section{Introduction}

In the Regge pole model the scattering amplitude
at large energies $\sqrt{s} $ and fixed
momentum transfers $\sqrt{-t}$ has the form~\cite{Grib1}
\beq
A_{Regge}^p(s,t)=\xi _p(t)\,s^{1+\omega _p(t)}\,\gamma
^2(t)\,,\,\,\xi_p(t)=e^{-i\pi \omega_p
(t)}-p\,,
\eeq
where $p=\pm 1$ is the signature of the reggeon with the trajectory
$\omega _p(t)$. The Pomeron is the Regge pole
of the $t$-channel partial waves $f_\omega (t)$ with vacuum quantum
numbers and the
positive signature describing an approximately constant behaviour of
total cross-sections for the hadron-hadron scattering.
S. Mandelstam demonstrated,
that the Regge poles generate  cut singularities in the
$\omega$-plane~\cite{Mand}. These singularities appear as a result of the analytic
continuation of the multi-particle unitarity condition~\cite{Grib2}.
They correspond to scattering states of the reggeons. Using the locality of the
reggeon interactions in the rapidity space, V. Gribov constructed an effective
(2+1) Pomeron field
model~\cite{Grib3}.

On the other hand, it was discovered, that in some field
theories the elementary particles become reggeons after taking into account
radiative corrections. The simplest example of this phenomenon is
the electron reggeization in quantum electrodynamics with a massive
photon~\cite{GGLMZ}. Using the counting rules suggested in Ref.~\cite{Mand1}
the vector boson reggeization in the gauge models with the Higgs mechanism
was also established~\cite{GST}.

In the leading logarithmic approximation (LLA)
the scattering amplitude at high energies in QCD has the Regge form~\cite{BFKL}
\beq
M_{AB}^{A^{\prime }B^{\prime }}(s,t)=M_{AB}^{A^{\prime }B^{\prime
}}(s,t)|_{Born}\,s^{\omega (t)}\,,
\eeq
where $M_{Born}$ is the Born amplitude and the gluon Regge trajectory is given below
\beq
\omega (-|q|^2)=-\frac{\alpha _s N_c}{4\pi ^2}\,
\int d^2k\,\frac{|q|^2}{|k|^2|q-k|^2}
 \approx
-\frac{\alpha _sN_c}{2\pi }\,\ln \frac{|q^2|}{\lambda ^2}\,,
\eeq
where $\lambda$ is a gluon mass, introduced for the infrared regularization.

In the multi-Regge kinematics, where the pair energies $\sqrt{s_k}$ of
the produced gluons are large in comparison with momentum transfers $|q_i|$
the production amplitudes
in LLA are constructed from products of the Regge factors
$s_k^{\omega (t_k)}$ and effective reggeon-reggeon-gluon vertices
$C_\mu (q_r, q_{r+1})$~\cite{BFKL}. The amplitudes
satisfy the Steinmann relations and the $s$-channel unitarity
leading to bootstrap equations~\cite{BFKL}.

The knowledge of $M_{2\rightarrow 2+n}$ allows us to construct
the BFKL equation for the Pomeron
wave
function~\cite{BFKL}
\beq
E\,\Psi (\vec{\rho}_{1},\vec{\rho}_{2})=H_{12}\,\Psi (\vec{\rho}_{1},\vec{%
\rho}_{2})\;,\,\,\sigma _t \sim s^\Delta \,,\,\,\Delta =-\frac{\alpha
_{s}N_{c}}{2\pi }\,E_0\,.
\label{BFKLeq}
\eeq
Here $H_{12}$ is the BFKL Hamiltonian and $\Delta$ is the Pomeron intercept.
The operator $H_{12}$ has the property of the holomorphic separability~\cite{int1}
and is invariant under the M\"obius
transformations~\cite{moeb}
\beq
{\large \rho _k \rightarrow \frac{a\rho _k+b}{c\rho _k+d}}\,.
\eeq
The generalization of eq. (\ref{BFKLeq}) to a composite state of several gluons~\cite{BKP} in the multi-color QCD
leads to an integrable
XXX model~\cite{int} having a duality symmetry~\cite{dual}.

The next-to-leading correction to the BFKL kernel in QCD is also calculated~\cite{FL}.
Its eigenvalue contains non-analitic
terms
proportional to $\delta _{n,0}$ and $\delta_{n,2}$, where $n$ is the conformal spin of the M\"{o}bius group.
But in the case of the
$N=4$ extended supersymmetric gauge model these Kronecker symbols are canceled
leading to an expression having the properties of the hermitian
separability~\cite{trajN4} and maximal transcendentality~\cite{KL}.
The last property allowed to calculate
the anomalous dimensions of twist-two operators up to three loops~\cite{KLOV}.
It turns out, that evolution equations for the so-called
quasi-partonic operators are integrable in $N=4$ SUSY at the multi-color limit~\cite{integrQP}.
The $N=4$  four-dimensional conformal field theory due to the Maldacena guess is
equivalent to the
superstring living
on the anti-de-Sitter 10-dimensional space~\cite{Malda, GKP, W}.
Therefore the Pomeron in N=4
SUSY is equivalent
to the reggeized graviton in this space. This equivalence gives a possibility
to calculate the intercept of the BFKL Pomeron at large coupling
constants~\cite{KLOV, BPST}
\beq
j=2-\Delta \,,\,\,
\Delta =\frac{1}{2\pi} \,\hat{a}^{-1/2}\,,\,\,a=\frac{g^2N_c}{16 \pi ^2}\,.
\eeq
The duality between the BFKL Pomeron and the reggeized graviton means, that the
Pomeron calculus  could be described in a framework of the
approach based on the effective action for the reggeized gravitons. It is
one of the reasons, why we investigate in this paper the gravity at high energies.

To begin with, let us remind the effective field theory for reggeized
gluons~\cite{eff}.
The corresponding effective action is local in the rapidity space
\beq
y=\frac{1}{2}\ln \frac{\epsilon _{k}+|k|}{\epsilon _{k}-|k|}\,,\,\,
|y-y_{0}|<\eta \,,\,\,\eta <<\ln \,s \,.
\eeq
We introduce the anti-hermitian fields $v_\mu$ for the usual gluons
and the gauge invariant fields $A_\pm$  describing the production and annihilation
of the reggeized gluons
\beq
v_{\mu }(x)=-iT^{a}v_{\mu }^{a}(x)\,,\,\,A_{\pm }(x)=-iT^{a}A_{\pm }^{a}(x)
\,,\,\delta A_{\pm }(x)=0\,,
\eeq
where $T^a$ are generators of the gauge group in the adjoint representation.
The fields $A_\pm$ satisfy the kinematical constraints
\beq
\partial _-A_+=\partial _+A_-=0\,.
\label{constraint}
\eeq

The effective action for a cluster of particles with approximately equal
rapidities has the form
\beq
S=\int d^{4}x\left( L_{QCD}+Tr (V_+\partial _\mu ^2A_-+V_-\partial _\mu
^2A_+)\right) \,,
\eeq
where $L_{QCD}$ is the usual QCD action and the effective currents $V_\pm$ are
given
below
\beq
V_+=-\frac{1}{g}\partial _{+}\,P\exp \left( -\frac{g}{2}
\int_{-\infty }^{x^{+}}v_{+}(x^{\prime })d(x^{\prime })^{+}\right)
=v_+-gv_+\frac{1}{\partial_+}v_++...\,.
\eeq
The Feynman rules for this action are derived in ref.~\cite{last}. The effective
action approach
gives a possibility to construct various reggeon vertices needed to calculate
NLO and NNLO corrections to the BFKL kernel.

Another application of the effective action for the gauge models  is a verification of the
BDS ansatz~\cite{BDS} for the inelastic amplitudes in
$N=4$ SUSY. It was shown~\cite{BLS1, BLS2}, that the BDS amplitude $M^{BDS}_{2\rightarrow 4}$
should be
multiplied by the factor
containing the contribution of the Mandelstam cut in LLA. In the two-loop approximation
this factor can be found
also from properties of analyticity
and factorization~\cite{Lip10}
or directly from recently obtained exact result for $M_{2\rightarrow
4}$~\cite{LipPryg}. In a general case the wave function for the Mandelstam singularity  satisfies
the Schr\"{o}dinger
equation for an open integrable Heisenberg spin chain~\cite{Intopen}.

Below we apply the approach based on the effective action for reggeons to the case
of the high energy gravity. The graviton reggeization was established initially with the
use of the $t$-channel unitarity~\cite{GS}. Later it was shown~\cite{Lip1}, that in LLA
the graviton Regge trajectory is finite in the ultraviolet region only in the $N=4$ supergravity. At
other gravity models the corresponding $t$-channel partial wave $f_j(t)$ has a Regge cut
singularity corresponding to the double-logarithmic asymptotics~\cite{Lip1}. Also some effective vertices
for reggeized
graviton interactions were calculated~\cite{Lip2}. These
results were verified by the authors of Ref.~\cite{BAC} in their study of the gravity at Planckian
energies. An effective field theory for the $S$-matrix in gravity with the multi-Regge unitarity
was constructed in Ref.~\cite{Lip3}, which allowed to investigate the gravitational collapse
at the high energy particle scattering~\cite{ACV}. The new results on the high energy scattering
in gravity and supergravity and related references can be found in the paper~\cite{GSA}.

\section{Reggeon fields in gravity}
It is natural to construct the theory of high energy processes in gravity in terms of the
reggeized gravitons and their interactions, because in this case the scattering amplitudes will
satisfy automatically the $t$-channel unitarity. The $S$-channel unitarity will be
incorporated in the reggeon vertices. In particular, the so-called bootstrap relations in QCD are consequences
of the multi-particle $S$-channel unitarity. We begin with the introduction of the fields
describing the usual and reggeized gravitons.

The Hilbert-Einstein action for gravity has the
form~\cite{Fock}
\beq
S=S_{grav}+S_{m}\,,
\eeq
where $S_{grav}$ and $S_{m}$ describe
the interaction of the gravity field $g_{\mu \nu} (x)$ and the matter fields,
respectively. Both contributions
are invariant under the general coordinate transformation. For the
metric tensor $g_{\mu \nu}$, entering in the invariant interval
\beq
(ds)^2=\sum _{\mu \nu}g_{\mu \nu}\,dx^{\mu}\,dx^{\nu}\,,
\eeq
this transformation in the infinitesimal form is given below
\beq
\delta g_{\mu \nu}(x)=D_{\mu}\,\chi _{\nu }(x)+D_{\nu}\,\chi _{\mu }(x)\,,
\eeq
where $\chi _{\sigma }(x)$ is a small local parameter
and $D_{\sigma}$ is the covariant derivative defined below
\beq
D_{\mu}\,\chi _{\nu }(x)=\partial _{\mu} \chi _{\nu }(x)-\Gamma ^{\rho}_{\mu \nu}\,
\chi _{\rho}(x)\,.
\eeq
Here $\partial _{\mu}$ is the partial derivative in $x^{\mu}$ and
$\Gamma ^{\rho}_{\mu \nu}$ is the Christoffel symbol
\beq
\Gamma ^{\rho}_{\mu \nu}=\frac{1}{2}\,g^{\rho \sigma}\,\left(
\partial_\mu g_{\sigma \nu}+\partial_\nu g_{\mu \sigma}
-\partial_\sigma g_{\mu \nu}\right)\,.
\eeq
Note, that in the Minkowski space
the corresponding invariant interval is
\beq
(ds)^2=\eta _{\mu \nu}\,dx^\mu\,dx^\nu \,,
\eeq
where the Lorentz tensor $\eta ^{\mu \nu}$ has the signature ($+---$).

The action for the pure gravity can be written as follows
\beq
S_{grav}=-\frac{1}{2\kappa}
\,\int d^4x\,\sqrt{-g}\,R\,,
\eeq
where
\beq
g=Det (g_{\mu \nu})
\eeq
and the Einstein parameter $\kappa$ is proportional to the Newton constant $\gamma$
\beq
\kappa =8\,\pi \,\gamma\,.
\eeq
The scalar curvature $R$ is related to the Riemann tensor by the contraction of indices
\beq
R=R_{\mu \nu}\,g^{\mu \nu}\,.
\eeq
In turn, $R_{\mu \nu}$ is expressed in terms of the Riemann tensor of the fourth rank
\beq
R_{\mu \nu}=R^\sigma _{\mu ,\sigma \nu}\,,\,\,R^\sigma _{\mu ,\alpha \beta}=
\partial _\beta \Gamma ^\sigma_{\mu \alpha }-
\partial _\alpha \Gamma ^\sigma_{\mu \beta }+\Gamma ^\rho_{\mu \alpha }\,
\Gamma ^\sigma_{\rho \beta }-\Gamma ^\rho_{\mu \beta }\,\Gamma ^\sigma_{\rho \alpha }\,.
\eeq

The matter action  can be written as an integral from the Lagrangian $L_{m}$
\beq
S_{m}=\int d^4x\,\sqrt{-g}\,L_{m}\,.
\eeq
Its variation in the metric tensor is expressed in terms of the energy-momentum tensor
$T^{\mu \nu}$
\beq
\delta S_{m}=\frac{1}{2}\int d^4 x\,\sqrt{-g}\,\delta g_{\mu \nu}\,T^{\mu \nu}\,,
\eeq
which is conserved
\beq
D_\mu \,T^{\mu \nu}=0\,.
\eeq
Performing also the variation of the gravity action
\beq
\delta S_{grav}=\frac{1}{2\kappa}\int d^4x\,\sqrt{-g}\,\delta g_{\mu \nu}\,G^{\mu \nu}
\,,\,\,G^{\mu \nu}\equiv R^{\mu \nu}-\frac{1}{2}\,g^{\mu \nu}R\,,
\eeq
we obtain from the stationarity condition for the total action $\delta S=0$ the Einstein equations
\beq
G^{\mu \nu}=-\kappa T^{\mu \nu}
\eeq
Due to the general covariance of the action $S_{grav}$ the left hand side of this
equality satisfies the identity
\beq
D_\mu G^{\mu \nu}=0\,,
\eeq
which is compatible with the conservation of the energy-momentum tensor $T^{\mu \nu}$.

We want to construct the interaction of the Reggeized gravitons and usual gravitons
to describe quasi-multi-Regge processes at high energies. In these processes intermediate
particles are produced
in clusters with fix invariant masses $m_r$. The particles in each cluster have
approximately the same rapidities $y\approx y_r$ in the interval
\beq
|y-y_r|<\eta \,,
\eeq
 where the intermediate parameter $\eta$ is assumed to satisfy the inequalities
\beq
\ln s \gg \eta \gg 1\,.
\eeq
The relative rapidities of different clusters produced in the multi-Regge kinematics are
considered to be large
\beq
\ln s \gg y_{r}-y_{r-1} \gg \eta \,.
\eeq

The interaction between the clusters is performed through an exchange of the reggeized
gravitons described by two additional fields $A^{++}$ and $A^{--}$. These new fields
correspond to the reggeon emission and absorption
in the crossing channel. In our quasi-multi-Regge kinematics corresponding to the strongly
ordered rapidities they satisfy the following
kinematical constraints (cf. (\ref{constraint}))
\beq
\partial _+A^{++}=0\,,\,\,\partial _{-}A^{--}=0\,,
\label{constraint}
\eeq
where
\beq
\partial _\pm =n_\pm^\sigma \partial _\sigma \,,\,\,n_\pm ^2=0\,,\,\,n_+n_-=1
\eeq
and the light-cone vectors $n_\pm$ are expressed in terms of momenta $p_A,p_B$ of colliding
particles
\beq
n_+=p_A\,\sqrt{\frac{2}{s}}\,,\,\,n_-=p_B\,\sqrt{\frac{2}{s}}\,.
\eeq
The above constraints for $A^{\pm}$ follow from the fact that  the Sudakov
variables $\alpha _r,\,\beta_r$ for the
produced cluster momentum
\beq
k_r=\beta _rp_A+\alpha_rp_B+k_{r\perp}\,,\,\,k_r^2=s\alpha_r\beta_r-
{\bf k}_{r\perp}^2=m_r^2
\eeq
are strongly ordered
\beq
1\gg \beta_1\gg\beta_{2}\gg...\gg \beta _n\,,\,\,
1\gg \alpha_n\gg\alpha_{n-1}\gg...\gg \beta _1\,,\,\,s\alpha_r\beta_r
\sim {\bf k}_{r\perp} ^2\sim m^2_r\,.
\eeq

We derive in next sections the effective action for high energy processes in
gravity. This action describes the interaction of gravitons inside each cluster with
neighboring reggeons having approximately the same rapidities.
In this case apart from the usual
Hilbert-Einstein action $S_{grav}$ one should introduce an additional contribution
$\Delta S$ containing the linear combination of the reggeon fields $A^{++}$ and $A^{--}$
considered them as
external sources
\beq
S_{eff}=S_{grav}+\Delta S\,.
\eeq
It is well known, that the reggeon describes a family of particles with different spins and masses
lying on the Regge
trajectory. The reggeized graviton can be viewed as a natural generalization
of the usual massless
graviton with the spin $j=2$. Therefore it is natural to consider the
functions $A^{++}$ and $A^{--}$ as fields invariant under general covariant
transformations
\beq
\delta A^{++}=\delta A^{--}=0
\eeq
with the corresponding local parameters $\chi$ decreasing at large $x$. Of course, $A^{++}$ and
$A^{--}$ are transformed under the global Poincare group. The induced contribution
$\Delta S$ can be
written in the form
\beq
\Delta S =-\frac{1}{2\kappa}\,\int d^4x\,\left(j_{++}\,\frac{\partial^2 _\mu A^{++}}{2}
+j_{--}\,\frac{\partial^{2}_\mu A^{--}}{2}
\right)\,.
\eeq
Here the Laplacian operators $\partial _\mu^2$ are introduced to avoid simultaneous
singularities in the overlapping
direct and crossing channels because they cancel neighboring reggeon propagators $1/q^2$.
These propagators appear due to a kinetic term bilinear in the fields $A^{\pm \pm}$.
As it is shown below, due to general covariance the
currents $j_{++}$ and $j_{--}$ contain the non-local operators $\partial ^{-1}_+$
and $\partial ^{-1}_-$ which should be interpreted as
propagators
of the particles in other clusters emitting the gravitons into
the given rapidity interval $\eta$.

In the perturbation theory the currents $j_{++},\,j_{--}$
contain contributions linear in the metric tensor $g_{\mu \nu}$
\[
j_{++}\approx g_{++}+...\,,\,\,\,j_{--}\approx g_{--}+...\,.
\]
The corresponding solution of
the Einstein equation for $g^{\mu \nu}$ in the external fields $A^{++}$ and $A^{--}$ will
have the form
\beq
g^{\mu \nu}=\eta ^{\mu \nu}+\delta ^\mu_+\delta ^\nu_+\, A^{++}+
\delta ^\mu_-\delta ^\nu_-\, A^{--}
+O(A ^2)\,.
\eeq
Therefore the reggeon fields $A^{++}$ and $A^{--}$ can be considered as classical components
of the gravity field.

The induced term $\Delta S$ should be invariant under general coordinate
transformations providing that $A^{\pm \pm}$ satisfy the kinematical constraints (\ref{constraint}).
As it was argued above, the
currents $j_{++}$ and $j_{--}$ describe the graviton emission  into the given interval of
rapidities from other clusters having different rapidities. In an accordance with the condition
$\partial _\pm A^{\pm \pm }=0$
two neighboring  reggeons for the cluster $r$ have the momentum components $k^\pm _r$
which are transferred almost completely to the
particles in the  cluster.
These momenta are shared by the particles in other clusters
with higher values of $k^\pm$. Because the currents $j_{\pm }$ are universal, for their calculation
one can consider an arbitrary process in the external field having particles
with the larger components $k^\pm $ of their momenta.

\section{General covariance of the effective action}
To calculate the effective currents $j_{++}$ and $j_{--}$ we use their invariance under
the general coordinate transformations up to the total derivatives in $x^+$ and $x^-$,
taking into account the fact that the fields $A^{++}$ and $A^{--}$
are invariant under these transformations and satisfy additional constraints
\beq
\partial _+A^{++}=\partial _{-}A^{--}=0\,.
\eeq
As a gravitational field we chose the tensor $h_{\mu \nu}$ in the following decomposition of the covariant
metric
tensor
\beq
g_{\mu \nu}=\eta _{\mu \nu}+h_{\mu \nu}\,.
\eeq
The components of the contravariant metric tensor $g^{\mu \nu}$ can be found from the linear equation
\beq
g^{\mu \sigma}\,g_{\sigma \nu} =\delta _{\nu}^{\mu}\,.
\eeq
and are obtained by the perturbation expansion
\beq
g^{\mu \nu}=\eta ^{\mu \nu}-h_{\mu \nu}+h_{\mu \rho}h_{\rho \nu}-h_{\mu \rho}h_{\rho \delta }h_{\delta \nu}
+...\,.
\eeq
Note, that for the tensor $h_{\mu \nu}$ and its derivatives we shall use only lower components implying
the Minkowski signature in summation over the repeated indices.
The effective currents $j_{++}$ and $j_{--}$ can be calculated in the perturbation series over the
tensor components $h_{\mu \nu}$. For example, $j_{++}$ can be presented as follows
\beq
j_{++}=h_{++}+P^{(2)}_{++}(h)+P_{++}^{(3)}(h)+...\,,
\eeq
where the polynomials $P^{(k)}_{++}$ can contain derivatives $\partial _\sigma$ and
integral operators $1/\partial _{+}$ acting on the fields $h$. Furthermore, generally we differ the components $h_{++},\,
h_{\sigma +}$ and $h_{\rho \sigma}$.
The corresponding set of recurrent equations for the homogeneous polynomials $P_{++}^{(n)}$ are obtained from the general
covariance of the induced action using  the infinitesimal transformations with parameters
$\chi _\rho$ and $\chi _+$
\beq
\frac{\delta P_{++}^{(n)}}{\delta h_{\rho \sigma}}\,2\,\partial _\sigma \chi _\rho
+\frac{\delta P_{++}^{(n)}}{\delta h_{\rho  +}}\,\partial _+\chi _\rho  =\sum _{k=1}^{n-1}
\frac{\delta P_{++}^{(k)}}{\delta h_{\rho \sigma}}\,2\Gamma _{\rho \sigma }^\nu \chi _\nu \,,
\eeq
\beq
\frac{\delta P_{++}^{(n)}}{\delta h_{+ +}}\,2\,\partial _+ \chi _+
+\frac{\delta P_{++}^{(n)}}{\delta h_{\rho  +}}\,\partial _\rho\chi _+ =0\,.
\eeq
In the right hand side of the first equation one should leave only the terms of the order $n$ in
the perturbation series over $h_{\rho \sigma}$ using the above expansion for
$g^{\mu \nu}$. It is implied also  in these equations, that after the differentiation over $h_{\rho \sigma}, h_{\rho +}$
and $h_{++}$ the corresponding tensor
components in $P_{++}$ should be replaced by the subsequent factors.

The second equation can be easily solved. Namely, $P^{(n)}_{++}$ should contain the dependence from $h_{\sigma +}$
and $h_{+ +}$ only
in the form of the following linear combination
\beq
X_{\sigma +}=X_{+\sigma }=h_{\sigma +}-\frac{1}{2}\,\frac{\partial _\sigma}{\partial _+}\,h_{++}\,.
\label{Xsigma}
\eeq
Moreover, we can add to the solution of the first equation an arbitrary function of another linear combination
\beq
Z_{\rho \sigma}=h_{\rho \sigma}-2\frac{\partial _\rho}{\partial _+}\,h_{\sigma +}\,.
\eeq
It is convenient to introduce two independent variables: $X_{\sigma +}$ and
\beq
Y_{\rho \sigma}=h_{\rho \sigma}-2\frac{\partial _\rho}{\partial _+}\,h_{\sigma +}+
\frac{\partial_\rho \partial _\sigma}{\partial _+^2}\,h_{++}=h_{\rho \sigma}-
2\frac{\partial _\rho}{\partial _+}\,X_{\sigma +}\,.
\eeq
Then the left hand sides of the above equations do not contain the derivative in $Y_{\rho \sigma}$ and
the first equation can be written as follows
\beq
\frac{\partial P_{++}^{(n)}}{\partial X_{\sigma +}}\,\partial _+\chi _\sigma=
\sum _{k=1}^{n-1}
\frac{\delta P_{++}^{(k)}}{\delta h_{\rho \sigma}}\,2\Gamma _{\rho \sigma }^\nu \chi _\nu \,.
\eeq
Here the right hand side should be expressed in terms of the variables $X_{\sigma +}$ and $Y_{\rho \sigma}$.
In particular using
\beq
P_{++}^{(1)}=h_{++}
\eeq
for $P_{++}^{(2)}$ one can obtain the equation
\beq
\frac{\partial P_{++}^{(2)}}{\partial X_{\sigma +}}\,\partial _+\chi _\sigma=
-2\,X_{\sigma +}\, \partial _+\chi _\sigma \,,
\label{eqseqord}
\eeq
where the following relation was used
\beq
\delta h_{++}=2\partial _+\chi _+-2\chi ^\sigma \partial _+\,X_{\sigma +}
\eeq
with the subsequent integration over $x_+$ by parts.
Note, that  the first term $2\partial _+\chi _+$ in $\delta h_{++}$ gives a vanishing contribution
to the induced action $\Delta S$ in this order due to the
kinematical constraint
\beq
\partial _+A^{++}=0\,.
\eeq
Therefore from eq. (\ref{eqseqord}) we obtain
\beq
P_{++}^{(2)}=-X_{\sigma +}^2\,.
\eeq
To find $P_{++}$ in upper orders of the perturbation theory one should use the following relations
\beq
\delta h_{\sigma +}=\partial _\sigma \chi _++\partial _+ \chi _\sigma -
\chi ^\rho \left(\partial _+h_{\sigma \rho }+\partial _\sigma h_{+\rho }-\partial _\rho h_{+\sigma}\right)\,,
\eeq
\beq
\delta h_{\sigma \nu}=\partial _\sigma \chi _\nu +\partial _\nu \chi _\sigma -
\chi ^\rho \left(\partial _\nu h_{\sigma \rho }+\partial _\sigma h_{\nu \rho }-\partial _\rho h_{\nu \sigma}\right)\,.
\eeq
Thus, in the third order we obtain the equation
\beq
\frac{\partial P_{++}^{(3)}}{\partial X_{\sigma +}}\,\partial _+\chi _\sigma =
-2\chi _\sigma h_{\sigma \rho }\partial _+X_{\rho +}+2X_{\sigma +}\,\delta^{(2)} X_{\sigma +}\,,
\label{eqthirdord}
\eeq
where
\beq
\delta ^{(2)}X_{\sigma +}=-\chi ^\rho \left(\partial _+Y_{\rho \sigma }+\partial _\rho X_{\sigma +}+
\partial _\sigma X_{\rho +}\right)+\frac{\partial _\sigma}{\partial _+}\,\chi ^\rho
\partial _+X_{\rho +}
\label{delta(2)}
\eeq
enters in the infinitesimal transformation of $X_{\sigma +}$
\beq
\delta X_{\sigma +}=\partial _+\chi _\sigma +\delta ^{(2)}X_{\sigma +}\,.
\eeq
With the integration over $x^+$ by parts one can rewrite the equation (\ref{eqthirdord}) in the form
\[
\frac{\partial P_{++}^{(3)}}{\partial X_{\sigma +}}\,\partial _+\chi _\sigma =
(\partial _+\chi _\sigma )\,
X_{\rho +}\,(Y_{\rho \sigma}+Y_{\sigma \rho})
\]
\beq
-2\chi _\sigma \left(\frac{\partial _\sigma }{\partial _+}\,X_{\delta +} \right)\partial _+X_{\delta +}
-2(\partial _\sigma \chi _\delta )\left(\frac{1 }{\partial _+}\,X_{\sigma +} \right)\partial _+X_{\delta +}
-2(\partial _+^2 \chi _\delta )\left(\frac{\partial _\sigma }{\partial _+}\,
X_{\delta +} \right)\frac{1}{\partial _+}\,X_{\sigma +}\,.
\eeq
After that its solution can be easily found
\beq
P_{++}^{(3)}=X_{\sigma +}X_{\delta +}\,Y_{\sigma \delta}-2\,
\left(\frac{1}{\partial _+}\,X_{\sigma +}\right)
\left(\frac{\partial _\sigma }{\partial _+}\,X_{\delta +} \right)\partial _+X_{\delta +}\,.
\eeq
Integration by parts and using the expression for $Y_{\sigma \delta}$ one can simplify this result
\beq
P_{++}^{(3)}
=X_{\rho +}\,X_{\sigma +}\,
h_{\rho \sigma}-
X_{\sigma +}\,\frac{\partial_\sigma}{\partial _+}\, X_{\rho +}^2\,.
\eeq
Therefore we obtain for the effective currents $j_{++}$ and $j_{--}$ the perturbative expansion
\beq
j_{++}=h_{++}-X_{\sigma +} ^2+X_{\rho +}\,X_{\sigma +}h_{\rho \sigma}
-X_{\sigma +}\,\frac{\partial_\sigma}{\partial _+}\, X_{\rho +}^2+...\,,
\label{effpert1}
\eeq
\beq
j_{--}=h_{--}-X_{\sigma -} ^2+X_{\rho -}\,X_{\sigma -}h_{\rho \sigma}-
X_{\sigma -}\,\frac{\partial_\sigma}{\partial _-}\, X_{\rho -}^2+...\,.
\label{effpert2}
\eeq

One can verify, that with our accuracy these currents are transformed under the coordinate change as follows
\beq
\delta j_{\pm \pm }\approx 2\partial _{\pm} \left(\chi _\pm -h_{\pm \sigma }\chi _\sigma \right)
+\partial _\pm \, \left(\chi _\sigma -h_{\sigma \rho }\chi _\rho \right)\,
\frac{\partial_\sigma}{\partial _\pm} \left(h_{++}-X_{\sigma +} ^2\right)\,.
\eeq

It allows us to guess the law of transformations of $j_{\pm \pm}$ in a general case
\beq
\delta j_{\pm \pm }= 2\,\partial _{\pm} \,\chi ^\mp +
\partial _\pm \, \chi ^\sigma \,\frac{\partial_\sigma}{\partial _\pm}\,j_{\pm \pm }\,,
\,\,\chi ^\sigma =g^{\sigma \rho}\,\chi _\rho \,.
\eeq

\section{Rapidly moving scalar particle in a gravity field}
The general covariance conditions for the effective reggeon currents
\beq
(\delta h_{\rho \sigma })\,\frac{\delta }{\delta h_{\rho \sigma}}\,j_{++}=
(\delta h_{\rho \sigma })\,\frac{\delta }{\delta h_{\rho \sigma}}\,j_{--}=0
\eeq
are equivalent to equations of motion for the relativistic matter
propagating  in the
corresponding
gravitational field because the
effective action can be viewed as a backward reaction of the rapidly moving colliding particles on
the processes taking place at a given interval of rapidity. Due to the universality
of the action for its calculation one can consider an arbitrary type of the colliding matter.

Let us restrict ourselves to the
scattering of the high energy scalar particle off the gravitational field.
The action for the free massless scalar field $\phi$ in the gravitational background can be
written as follows
\beq
S_s=\int d^4x\,\sqrt{-g}\,\frac{1}{2}\,(\partial_\mu \phi )\,g^{\mu \nu}\,
(\partial_\nu \phi )
\,.
\eeq
The corresponding energy-momentum tensor is
\beq
T_{\mu \nu}=(\partial _\mu \phi )\,(\partial _\nu \phi ) -\frac{1}{2}\,g_{\mu \nu}\,g^{\rho \sigma }\,
(\partial _\rho \phi )\,(\partial _\sigma \phi )\,.
\eeq
We introduce the covariant d'Alambert  operator
\beq
\nabla^2=\frac{1}{\sqrt{-g}}\partial _\mu \,g^{\mu \nu}\sqrt{-g}\,\partial _\nu=D^\mu \partial _\mu \,,
\label{Dalamber}
\eeq
which is symmetric for the following scalar product of the fields
\beq
\int d^4x\,\sqrt{-g}\,\psi \,\nabla ^2\,\phi
=\int d^4x\,\sqrt{-g}\,\phi \,\nabla ^2\,\psi\,.
\eeq
The equations of motion for $\phi$ are
\beq
\nabla^2\phi =0\,.
\eeq
The energy-momentum tensor is conserved
\beq
D^\mu T_{\mu \nu}=0
\eeq
 due to the equations of motion.

One
can construct also the equation for the Green function $G(x,x')$ of the scalar particle
\beq
-\nabla ^2(x)\,G(x,x')=\delta^4 (x-x')\,.
\eeq
Its arguments can be interchanged with a similarity transformation
\beq
G(x,x')=\sqrt{-g(x')}\,G(x',x)\,\frac{1}{\sqrt{-g(x)}}\,.
\eeq
The variation of the Green function over the metric tensor can be written as follows
\[
\delta G(x,x')=\int d^4y \,G(x,y)\,\delta (\sqrt{-g}\nabla^2)\,G(y,x')
\]
\beq
=-
\int d^4y \,\frac{\partial G(x,y)}{\partial y^\mu}\,\delta (g^{\mu \nu} \sqrt{-g})\,
\frac{\partial G(y,x')}{\partial y^\nu}\,.
\eeq

Under the general covariant transformations which
can be written in the form
\beq
\delta (g^{\mu \nu} \sqrt{-g})=\sqrt{-g}\,\left(g^{\mu \sigma}D_\sigma \chi ^{\nu }+
g^{\mu \sigma}D_\sigma \chi ^{\nu }-g^{\mu \nu}D_\sigma \chi ^{\sigma}\right)\,.
\eeq
the Green function is transformed as follows
\beq
\delta G(x,x')=\chi ^\sigma (x)\,\frac{\partial}{\partial x^\sigma } G(x,x')+
\chi ^\sigma (x')\,\frac{\partial}{\partial x^{\prime \sigma} } G(x,x')\,.
\eeq

The corresponding $S$-matrix exists providing that at infinity the metric has the Minkowski form
\beq
\lim _{x\rightarrow \infty}g^{\mu \nu}= \eta ^{\mu \nu}\,.
\eeq
In this case the scattering amplitude $f(p,p')$ is expressed in terms of the matrix
element of
the Green function with amputated free propagators
\beq
f(p,p')\sim  <p'|\,\lim _{t\rightarrow \infty}\,\lim _{t'\rightarrow -\infty}\,
\partial _\sigma ^2\,G(x,x')\,
\partial _{\sigma '}^2\,
|p>\,,
\eeq
where the initial and final particles are on mass shell
\beq
p^2=p^{\prime 2}=0\,.
\eeq

The scattering amplitude is invariant under the general coordinate transformations,
because the infinitesimal parameter $\chi$ tends to zero at infinity. Note, however, that
generally the energy and momentum are not conserved.

For our purpose it is enough to find the Green function only at
high energies
\beq
p_\sigma \approx p'_\sigma \rightarrow \infty \,.
\eeq
For example let us consider the colliding particle with the momentum
\beq
p_A =n_+\,\sqrt{\frac{s}{2}}\,.
\eeq
In this case the wave functions $<p_A|$ and $<p_{A'}|$ are rapidly
oscillate and
we can write the covariant d'Alambert  operator in the equation for $G(x,x')$ as follows
\beq
\nabla ^2=\partial _\sigma ^2 +h^{--}\partial _-^2+2(\partial_\sigma h^{\sigma -})\partial _-+
\frac{1}{2}(\partial _+ h_{\rho \sigma})g^{\rho \sigma}\partial_-+
\frac{1}{2}h^{\mu -}(\partial _\mu h_{\rho \sigma})g^{\rho \sigma}\partial_-\,,
\eeq
where we introduced the notations
\beq
g^{\rho \sigma}=\eta _{\rho \sigma}+h^{\rho \sigma}\,,\,\,
h^{\mu -}=
-h_{\mu +}+h_{\mu \rho}h_{\rho +}+...\,,
\eeq
\beq
h^{--}=-h_{++}+h_{+\rho}h_{\rho _+}-h_{+\rho}h_{\rho \sigma}h_{\sigma _+}+...\,.
\eeq
We imply also the following decomposition of the usual Laplace operator
\beq
\partial _\sigma ^2=\partial _+\partial _-+\partial _\sigma \partial _\sigma \,.
\eeq

The closed expression for the induced current $j_{++}$ is given
below
\beq
j_{++}=\frac{\partial _\rho ^2}{\partial_-}\,\nabla ^{-2}\,
\frac{\partial _\sigma ^2}{\partial_-}\,,
\eeq
where it is implied, that the differential operators in the end of the expression act to the left
after their integration by parts. One can use also a semiclassical approximation for the Green function.
We shall return to the semiclassical approach in another form in subsequent
sections.

\section{"Eikonal" contribution to the effective action}
As it was mentioned above, in the perturbation theory the scalar particle in the intermediate
states is strongly virtual in an accordance with the fact, that
in our kinematics
the gravitons emitted from it belong to the clusters
with their rapidity  significantly different from the particle rapidity.
Therefore we
can expand its free
propagator as follows
\beq
-\frac{1}{\partial _\sigma ^2}\approx -\frac{1}{\partial _+ \partial _-}+
\frac{1}{\partial _+ \partial _-}
\partial^\perp _\sigma \partial^\perp _{ \sigma}
\,\frac{1}{\partial _+ \partial _-}\,.
\eeq
The leading terms $\sim h^{++}$ are canceled partly in the perturbation expansion
between
contributions of various
Feynman diagrams corresponding to a different ordering of the vertices $S_{int}$ in time,
because the eikonal term with intermediate particles on mass shell in our case is negligible.
To clarify this important fact we calculate here several terms of the expansion of scattering amplitude in the
Fourier transform $V(k)$ of the interaction term $h_{++}$. Omitting the normalization factors
and the vertices $V(k_i)$, the scattering amplitude for  the scalar particle
with the large momentum $p$ can be written in the second order of perturbation theory
as follows
\beq
A_2^{eik}=\frac{1}{(p+k_1)^2}+\frac{1}{(p+k_2)^2}=
\frac{(p+k_1+k_2)^2-2(k_1k_2)}{(p+k_1)^2(p+k_1+k_2)^2}\approx
-\frac{(k_1k_2)}{2(pk_1)\,(pk_2)}\,,
\eeq
where we used the reality requirement for the initial and final state particles
\beq
p^2=(p+k_1+k_2)^2=0
\eeq
and the condition of the strong virtuality for the particle in the intermediate states
\beq
2(pk_1)\sim 2(pk_2)\gg k_1^2\sim k_2^2\sim (k_1k_2)\,.
\eeq
In an analogous way one can obtain the following contributions from the eikonal diagrams in the third
\[
A_3^{eik}=\frac{1}{(p+k_1)^2}\frac{1}{(p+k_1+k_2)^2}+\frac{1}{(p+k_2)^2}\frac{1}{(p+k_1+k_2)^2}+
\frac{1}{(p+k_1)^2}\frac{1}{(p+k_1+k_3)^2}
\]
\[
+\frac{1}{(p+k_3)^2}\frac{1}{(p+k_1+k_3)^2}+
\frac{1}{(p+k_2)^2}\frac{1}{(p+k_2+k_3)^2}+\frac{1}{(p+k_3)^2}\frac{1}{(p+k_2+k_3)^2}
\]
\beq
\approx \frac{k_{3}(k_1+k_2)}{4pk_3\,p(k_1+k_2)}\,
\frac{k_{1}k_2}{pk_1\,pk_2}+
\frac{k_{2}(k_1+k_3)}{4pk_2\,p(k_1+k_3)}\,
\frac{k_{1}k_3}{pk_1\,pk_3}+\frac{k_{1}(k_2+k_3)}{4pk_1\,p(k_2+k_3)}\,
\frac{k_{2}k_3}{pk_2\,pk_3}
\eeq
and fourth order
\[
A_4^{eik} \approx -\frac{k_{1}k_2}{8\,pk_1\,pk_2}\left(\frac{k_{4}(k_1+k_2+k_3)}{pk_4\,p(k_1+k_2+k_3)}\,
\frac{k_{3}(k_1+k_2)}{pk_3\,p(k_1+k_2)}+\frac{k_{3}(k_1+k_2+k_4)}{pk_3\,p(k_1+k_2+k_4)}\,
\frac{k_{4}(k_1+k_2)}{pk_4\,p(k_1+k_2)}\right)
\]
\[
-\frac{k_{1}k_3}{8\,pk_1\,pk_3}\left(\frac{k_{4}(k_1+k_2+k_3)}{pk_4\,p(k_1+k_2+k_3)}\,
\frac{k_{2}(k_1+k_3)}{pk_2\,p(k_1+k_3)}+\frac{k_{2}(k_1+k_3+k_4)}{pk_2\,p(k_1+k_3+k_4)}\,
\frac{k_{4}(k_1+k_3)}{pk_4\,p(k_1+k_3)}\right)
\]
\[
-\frac{k_{1}k_4}{8\,pk_1\,pk_4}\left(\frac{k_{3}(k_1+k_2+k_4)}{pk_3\,p(k_1+k_2+k_4)}\,
\frac{k_{2}(k_1+k_4)}{pk_2\,p(k_1+k_4)}+\frac{k_{2}(k_1+k_3+k_4)}{pk_2\,p(k_1+k_3+k_4)}\,
\frac{k_{3}(k_1+k_4)}{pk_3\,p(k_1+k_4)}\right)
\]
\[
-\frac{k_{2}k_3}{8\,pk_2\,pk_3}\left(\frac{k_{4}(k_1+k_2+k_3)}{pk_4\,p(k_1+k_2+k_3)}\,
\frac{k_{1}(k_2+k_3)}{pk_1\,p(k_2+k_3)}+\frac{k_{1}(k_2+k_3+k_4)}{pk_1\,p(k_2+k_3+k_4)}\,
\frac{k_{4}(k_2+k_3)}{pk_4\,p(k_2+k_3)}\right)
\]
\[
-\frac{k_{2}k_4}{8\,pk_2\,pk_4}\left(\frac{k_{3}(k_1+k_2+k_4)}{pk_3\,p(k_1+k_2+k_4)}\,
\frac{k_{1}(k_2+k_4)}{pk_1\,p(k_2+k_4)}+\frac{k_{1}(k_2+k_3+k_4)}{pk_1\,p(k_2+k_3+k_4)}\,
\frac{k_{3}(k_2+k_4)}{pk_3\,p(k_2+k_4)}\right)
\]
\[
-\frac{k_{3}k_4}{8\,pk_3\,pk_4}\left(\frac{k_{2}(k_1+k_3+k_4)}{pk_2\,p(k_1+k_3+k_4)}\,
\frac{k_{1}(k_3+k_4)}{pk_1\,p(k_3+k_4)}+\frac{k_{1}(k_2+k_3+k_4)}{pk_1\,p(k_2+k_3+k_4)}\,
\frac{k_{2}(k_3+k_4)}{pk_2\,p(k_3+k_4)}\right)
\]
\[
-\frac{(k_3+k_4)(k_1+k_2)}{8p(k_3+k_4)\,p(k_1+k_2)}\,
\frac{k_{1}k_2}{pk_1\,pk_2}\,
\frac{k_{3}k_4}{pk_3\,pk_4}-\frac{(k_2+k_4)(k_1+k_3)}{8p(k_2+k_4)\,p(k_1+k_3)}\,
\frac{k_{1}k_3}{pk_1\,pk_3}\,
\frac{k_{2}k_4}{pk_2\,pk_4}
\]
\beq
-\frac{(k_2+k_3)(k_1+k_4)}{8p(k_2+k_3)\,p(k_1+k_4)}\,
\frac{k_{1}k_4}{pk_1\,pk_4}\,
\frac{k_{2}k_3}{pk_2\,pk_3}\,.
\eeq
The corresponding  "eikonal" terms indeed appear  in the effective currents
\beq
j^{eik}_{++}\approx g_{++}-X_{\sigma +}^2-X_{\sigma +}\frac{\partial _\sigma}{\partial _+}\,X_{\rho +}^2-
X_{\mu +}\frac{\partial _\mu}{\partial _+}X_{\sigma +}\frac{\partial _\sigma}{\partial _+}\,X_{\rho +}^2
-\frac{1}{4}\,\left(\frac{\partial _\sigma}{\partial _+}\,X_{\rho +}^2\right)^2+...\,,
\label{eikcur1}
\eeq
\beq
j^{eik}_{--}\approx g_{--}-X_{\sigma -}^2-X_{\sigma -}\frac{\partial _\sigma}{\partial _-}\,X_{\rho -}^2-
X_{\mu -}\frac{\partial _\mu}{\partial _-}X_{\sigma -}\frac{\partial _\sigma}{\partial _-}\,X_{\rho -}^2
-\frac{1}{4}\,\left(\frac{\partial _\sigma}{\partial _-}\,X_{\rho -}^2\right)^2+...\,,
\label{eikcur2}
\eeq
where we took into account, that due to the general covariance
the light-cone components $h_{++}$ and $h_{--}$ can enter in the final
expressions only inside the tensors $X_{\sigma +}$ and $X_{\sigma -}$, respectively.

Looking at these expressions and comparing them with the above perturbative contributions obtained from general covariance
considerations one can formulate the hypothesis, that the complete  result for the generally invariant currents
is obtained from the "eikonal" expression by its "covariantization" corresponding to the substitution of
the Minkowski tensor $\eta ^{\mu \nu }$ everywhere by the world metric tensor:
\beq
\eta ^{\mu \nu}\rightarrow g^{\mu \nu}\,.
\eeq
This hypothesis  leads to the following result in the perturbation theory
\[
j_{++}= h_{++}-X_{\sigma +}g^{\sigma \rho}X_{\rho +}-
X_{\sigma +}g^{\sigma \rho}\frac{\partial _\rho}{\partial _+}\,X_{\mu +}g^{\mu \nu}X_{\nu +}
\]
\beq
-X_{\alpha +}g^{\alpha \beta}\frac{\partial _\beta}{\partial _+}\,X_{\sigma +}g^{\sigma \rho}
\frac{\partial _\rho}{\partial _+}\,X_{\mu +}g^{\mu \nu}X_{\nu +}
-\frac{g^{\sigma \rho}}{4}\,\left(\frac{\partial _\sigma}{\partial _+}\,X_{\mu +}g^{\mu \nu}X_{\nu +}\right)
\,\frac{\partial _\rho}{\partial _+}\,X_{\alpha +}g^{\alpha \beta}X_{\beta +}+...\,,
\eeq
\[
j_{--}= h_{--}-X_{\sigma -}g^{\sigma \rho}X_{\rho -}-
X_{\sigma -}g^{\sigma \rho}\frac{\partial _\rho}{\partial _-}\,X_{\mu -}g^{\mu \nu}X_{\nu -}
\]
\beq
-X_{\alpha -}g^{\alpha \beta}\frac{\partial _\beta}{\partial _-}\,X_{\sigma -}g^{\sigma \rho}
\frac{\partial _\rho}{\partial _-}\,X_{\mu -}g^{\mu \nu}X_{\nu -}
-\frac{g^{\sigma \rho}}{4}\,\left(\frac{\partial _\sigma}{\partial _-}\,X_{\mu -}g^{\mu \nu}X_{\nu -}\right)
\,\frac{\partial _\rho}{\partial _-}\,X_{\alpha -}g^{\alpha \beta}X_{\beta -}+...\,.
\eeq
Moreover, it allows to formulate a closed equation for the important "eikonal" contribution
\[
(\partial _+\chi _\rho )\,\frac{\delta}{\delta X_{\rho +}}\,j^{eik}_{++}
=
\left((\partial _\rho \chi _\sigma )+(\partial _\sigma \chi _\rho )\right)\,
\frac{\delta }{\delta \eta _{\rho \sigma }}j^{eic}_{++}
\]
\beq
-
\left(\chi _\rho (\partial _\rho X_{\sigma +})+(\partial _\sigma \chi _\rho )\,X_{\rho +}-
\left(\frac{\partial _\sigma}{\partial _+}\,(\partial _+\chi _\rho )\,X_{\rho _+}\right)\right)\,
\frac{\delta}{\delta X_{\sigma +}}\,j^{eik}_{++}\,,
\label{eqeik}
\eeq
where, as usual, the factors in front of derivatives should substitute in the same position
the corresponding variables
$X_{\sigma +}$ and $\eta _{\rho \sigma }$ removed by the differentiation. The first term
in the right hand side of this equation corresponds to the infinitesimal transformation
of $h_{\rho \sigma}$ in the lowest order of the perturbation theory.

For example, in the fourth order from this "eikonal" equation we derive the identity
\[
-(\partial _+\chi _\mu )\frac{\partial _\mu }{\partial _+}X_{\sigma +}\frac{\partial _\sigma }{\partial _+}\,X_{\rho +}^2
-X_{\mu +}\frac{\partial _\mu }{\partial _+}(\partial _+\chi _\sigma )\frac{\partial _\sigma }{\partial _+}\,X_{\rho
+}^2
\]
\[
-2\,X_{\mu +}\frac{\partial _\mu }{\partial _+}X_{\sigma +}\,\frac{\partial _\sigma }{\partial _+}X_{\rho +}
\partial _+\chi _\rho -\left(\frac{\partial _\sigma}{\partial _+}\,X_{\rho +}^2\right)
\frac{\partial _\sigma}{\partial _+}\,X_{\mu +}\partial_+\chi _\mu
\]
\[
\equiv -X_{\sigma +}\left((\partial _\sigma \chi _\rho )+(\partial _\rho \chi _{\sigma })\right)\frac{\partial _\rho }{\partial _+}X_{\mu +}^2
-2X_{\sigma +}\frac{\partial _\sigma  }{\partial _+}X_{\rho +} (\partial _\rho   \chi _\delta )X_{\delta +}
\]
\[
+\left(\chi _\rho (\partial _\rho X_{\sigma +})+(\partial _\sigma \chi _\rho )X_{\rho +}-\left(\frac{\partial _\sigma  }{\partial _+}
(\partial _+\chi _\rho )X_{\rho +}\right)\right)\frac{\partial _\sigma  }{\partial _+}X_{\mu +}^2
\]
\beq
+2\,X_{\mu +}\frac{\partial _\mu  }{\partial _+}X_{\sigma +}\,\left(\chi _\rho (\partial _\rho
X_{\sigma +})+(\partial _\sigma \chi _\rho )X_{\rho +})-\frac{\partial _\sigma  }{\partial _+}
(\partial _+\chi _\rho )X_{\rho +}\right)\,,
\eeq
which can be verified with integration over $x^+$ by parts.

In the fifth order one can obtain the relation
\[
(\partial _+\chi _\rho )\,\frac{\delta}{\delta X_{\rho +}}\,P^{eik \,(5)}_{++}=
-X_{\mu +} (\partial _\mu \chi _\nu )\frac{\partial _\nu }{\partial _+}\,X_{\sigma +}\frac{\partial _\sigma
}{\partial _+}X_{\rho +}^2
-X_{\mu +} \frac{\partial _\mu }{\partial _+}\,X_{\sigma +}\,(\partial _\sigma \chi _\delta )\frac{\partial _\delta
}{\partial _+}X_{\rho +}^2
\]
\[
-\frac{1}{2}\,(\partial _\sigma \chi _\rho )\left(\frac{\partial _\sigma
}{\partial _+}X_{\mu +}^2 \right)\left(\frac{\partial _\rho
}{\partial _+}X_{\nu +}^2 \right)+\left(\chi _\rho (\partial _\rho X_{\sigma +})-\left(\frac{\partial _\sigma  }{\partial _+}
(\partial _+\chi _\rho )X_{\rho +}\right)\right)\frac{\partial _\sigma}{\partial _+}\,X_{\delta +}
\frac{\partial _\delta}{\partial _+}X_{\mu +}^2
\]
\[
+X_{\mu +} \frac{\partial _\mu }{\partial _+}\,\left(\chi _\delta (\partial _\delta X_{\sigma +})-\left(\frac{\partial _\sigma  }{\partial _+}
(\partial _+\chi _\delta )X_{\delta +}\right)\right)\,\frac{\partial _\sigma
}{\partial _+}X_{\rho +}^2
\]
\beq
+\left(2X_{\mu +} \frac{\partial _\mu }{\partial _+}\,X_{\sigma +}+\left(\frac{\partial _\sigma  }{\partial _+} X^2_{\mu +}\right)
\right)\frac{\partial _\sigma  }{\partial _+} X_{\rho +}
\left(\chi _\delta (\partial _\delta X_{\rho +})-\frac{\partial _\rho  }{\partial _+}
(\partial _+\chi _\delta )X_{\delta +}\right)\,.
\eeq
It gives a possibility to calculate the corresponding "eikonal" contribution to $j_{++}$ in this order
\[
P^{eik \,(5)}_{++}=-X_{\nu +}\frac{\partial _\nu}{\partial _+}X_{\mu +}\frac{\partial _\mu}{\partial _+}
X_{\sigma +}\frac{\partial _\sigma}{\partial _+}\,X_{\rho +}^2
\]
\beq
-\frac{1}{4}\,X_{\nu +}\frac{\partial _\nu}{\partial _+}\,\left(\frac{\partial _\sigma}{\partial _+}\,X_{\rho +}^2\right)^2
-\frac{1}{2}\left(\frac{\partial _\sigma}{\partial _+}\,X_{\mu +}\frac{\partial _\mu}{\partial _+}\,X_{\rho +}^2\right)
\left(\frac{\partial _\sigma}{\partial _+}\,X_{\nu +}^2\right)\,.
\label{eikcur5}
\eeq
In the sixth order we obtain in a similar way
\[
P^{eik \,(6)}_{++}=-X_{\delta +}\frac{\partial _\delta}{\partial _+}X_{\nu +}\frac{\partial _\nu}{\partial _+}
X_{\mu +}\frac{\partial _\mu}{\partial _+}
X_{\sigma +}\frac{\partial _\sigma}{\partial _+}\,X_{\rho +}^2
\]
\[
-\frac{1}{4}\,X_{\delta +}\frac{\partial _\delta}{\partial _+}X_{\nu +}\frac{\partial _\nu}{\partial _+}\,
\left(\frac{\partial _\sigma}{\partial _+}\,X_{\rho +}^2\right)^2
-\frac{1}{2}\,X_{\delta +}\frac{\partial _\delta}{\partial _+}\left(\frac{\partial _\sigma}{\partial _+}\,
X_{\mu +}\frac{\partial _\mu}{\partial _+}\,X_{\rho +}^2\right)
\left(\frac{\partial _\sigma}{\partial _+}\,X_{\nu +}^2\right)
\]
\beq
-\frac{1}{4}\,
\left(\frac{\partial _\sigma}{\partial _+}\,
X_{\mu +}\frac{\partial _\mu}{\partial _+}\,X_{\rho +}^2\right)^2-\frac{1}{8}\,
\left(\frac{\partial _\sigma}{\partial _+}\,
\left(\frac{\partial _\delta}{\partial _+}\,X_{\rho +}^2\right)^2\right)
\left(\frac{\partial _\sigma}{\partial _+}\,X_{\nu +}^2\right)\,.
\eeq

To find a general structure for the currents $j_{++}^{eik}$ and $j_{--}^{eik}$ we should
investigate in a more accurate way
the recurrent relation following from the eikonal equation (\ref{eqeik}).

To begin with, one can use the following formulas (see (\ref{delta(2)}))
\[
2X_{\sigma _+}\delta ^{(2)}X_{\sigma +}-X_{\sigma +}\,X_{\rho _+}\,(\partial _\rho \chi _\sigma
+\partial _{\sigma }\chi _\rho)
\]
\beq
=-X_{\sigma +}\,\frac{\partial _\sigma }{\partial _+}\,2\,X_{\rho +}
(\partial _+\chi _\rho )-(\partial _+\chi _\sigma )\frac{\partial _\sigma }{\partial _+}X_{\rho +}^2
+\partial _+\chi _\sigma  \frac{\partial _\sigma }{\partial _+}\,X_{\rho +}^2\,.
\label{deltaX2}
\eeq
for the variation of the structure $X_{\sigma +} ^2$ present in the previous order. On the other hand, the sum of
the first two terms in the right hand side can be interpreted as the variation of the expression
\beq
-X_{\sigma +}\,\frac{\partial _\sigma }{\partial _+}\,X_{\rho +}^2
\eeq
appearing in the next order. The last term in (\ref{deltaX2})
gives a vanishing contribution in the second order. For higher orders it is multiplied
with two possible structure $X_\mu \partial _\mu /\partial _+$ or $\partial_\mu /\partial _+$.
The second structure is contracted
with the index
$\mu$ with the operator $\partial _\mu /\partial _+$ acting on another function. Let us consider these two
possibilities separately.

We obtain for the variation of the first structure
\[
\delta ^{(2)}X_{\mu +}\frac{\partial _\mu}{\partial_+ }-
X_{\mu +}\,(\partial _\nu \chi _\mu +\partial _{\mu }\chi _\nu )\frac{\partial _\nu}{\partial_+ }
\]
\beq
=\chi _\mu \partial _\mu  X_{\nu +}\frac{\partial _\nu }{\partial _+}-
X_ {\mu +} \partial _\mu  \chi _\nu \frac{\partial _\nu }{\partial _+}-
\left(\frac{\partial _\nu }{\partial _+}(\partial _+\chi _\mu )X_{\mu +}\right)
\frac{\partial _\nu }{\partial _+}\,.
\label{deltaX+}
\eeq
The second term in the right hand side cancels the contribution from the last term in
for the variation of $X_\sigma ^2$  (\ref{deltaX2}) due to the relation
\beq
-X_ {\mu +} \partial _\mu  \chi _\nu \frac{\partial _\nu }{\partial _+}X_{\sigma +} ^2+
X_{\mu +} \partial _\mu \chi _\sigma  \frac{\partial _\sigma }{\partial _+}\,X_{\rho +}^2=0
\eeq
The last term in  (\ref{deltaX+}) corresponds to the variation of the following structure in the next order
\beq
-\frac{1}{4}\left(\frac{\partial _\sigma }{\partial _+}\,X_{\rho +}^2\right)^2
\eeq
provided that the operator $X_{\mu +}\frac{\partial _\mu}{\partial_+ }$ was applied to $X_{\rho +}^2$. In other
cases we obtain from the last term the term canceling the variation of the contribution
\beq
-\frac{1}{2}\left(\frac{\partial _\nu}{\partial_+}\,X_{\mu +}^2\right)\frac{\partial _\nu}{\partial_+}\,.
\eeq
in the next order.
The first term in (\ref{deltaX+}) can be written as follows
\beq
-(\partial _+\chi _\mu )\frac{\partial _\mu}{\partial _+}
X_{\nu +}\frac{\partial _\nu}{\partial _+}+\partial _+\chi _\mu
\frac{\partial _\mu}{\partial _+}X_{\nu +}\frac{\partial _\nu}{\partial _+}\,.
\label{delXdel}
\eeq
Here the first contribution leads to the following structure in the next order
\beq
-X _{\mu +}\frac{\partial _\mu}{\partial _+} X_{\nu +}\frac{\partial _\nu}{\partial _+}
\eeq
and the second term vanishes provided it is not multiplied by  $X_\mu \partial _\mu /\partial _+$ or
$\partial_\mu /\partial _+$ contracted
by the index
$\mu$ with the operator $\partial _\mu /\partial _+$ acting on another function. In the these
two cases
we should repeat calculations performed above for the last term in the variation of $X_\sigma ^2$.

At last we consider  the variation of a product of the operators $\partial _\sigma /\partial _+$
in $h_{\sigma \rho}$ of
\beq
-(\partial _\sigma \chi _\rho +\partial _\rho \chi _\sigma )\,\frac{\partial _\sigma }{\partial _+}...
\frac{\partial _\rho }{\partial _+}=-\left(\partial _\rho \chi _\sigma
\frac{\partial _\sigma }{\partial _+}...\right)
\frac{\partial _\rho }{\partial _+}-\frac{\partial _\sigma }{\partial _+}...
\partial _\rho \chi _\rho
\frac{\partial _\rho }{\partial _+}+\chi _\mu \partial _\mu \left(\frac{\partial _\sigma }{\partial _+}...
\frac{\partial _\sigma }{\partial _+}...\right)
\,.
\eeq
Two first terms are canceled with the last terms in Eqs. (\ref{deltaX2}) and
(\ref{delXdel}). The last term can be written as follows
\beq
-(\partial _+\chi _\mu )\frac{\partial _\mu }{\partial _+}\left(\frac{\partial _\sigma }{\partial _+}...
\frac{\partial _\sigma }{\partial _+}...\right)+\partial _+\chi _\mu \frac{\partial _\mu }{\partial _+}
\left(\frac{\partial _\sigma }{\partial _+}...\frac{\partial _\sigma }{\partial _+}...\right)\,.
\label{gmunu}
\eeq
The first term here corresponds to the following structure in the next order
\beq
-X _{\mu +}\frac{\partial _\mu }{\partial _+}\left(\frac{\partial _\sigma }{\partial _+}...
\frac{\partial _\sigma }{\partial _+}...\right)\,.
\eeq
The second term is not zero only if it is multiplied by  $X_\nu \partial _\nu /\partial _+$ or
$\partial_\nu /\partial _+$ contracted
by the index
$\nu$ with the operator $\partial _\nu /\partial _+$ acting on another function. In these cases
again we should perform calculations similar to that with the last term in eq. (\ref{deltaX2}).
Thus, after cancelation of some terms in the variation of $j_{++}$ in the previous order
we can obtain the result in next order, using the above substitutions. 

Even more,
one can write the following representation for the effective "eikonal" currents in an arbitrary order
\beq
j_{++}^{eik}=h_{++}-\partial_+J_{+}^{eik}\,,\,\,j_{--}=h_{--}-\partial _-J_{-}^{eik}\,.
\eeq
The above perturbative calculations allow to construct the following "fan" equations
for  
the quantities $J_{\pm }^{eik}$ 
\beq
\left(\partial _\pm-X_{\sigma \pm }\partial_\sigma\right)\,J_\pm ^{eik}=X_{\mu \pm }^2 +\frac{1}{4}
\left(\partial _\rho J_\pm ^{eik}\right)^2\,.
\label{eqfan}
\eeq
The solutions of these equations should have the following transformation properties following
from the general coordinate invariance of $j_{\pm \pm}$ 
\beq
\delta J_\pm ^{eik} =-\frac{2}{\partial _\pm}\,\chi _\sigma \,\partial _\pm X_{\sigma \pm}=
\frac{2}{\partial _\pm}\,X_{\sigma \pm}\,\partial _\pm \chi _\sigma \,.
\eeq
One can verify, that indeed these properties are compatible with the transformations
of various operators entering in the "fan" equations
\beq
\delta \left(X_{\sigma \pm} \partial _\sigma \right)=(\partial _\pm \chi _\sigma  )
\partial _\sigma -\left(\frac{\partial _\sigma}{\partial _\pm }(\partial _\pm \chi _\mu )
X_{\mu \pm} \right)\partial _\sigma
-(\partial _\pm \chi _\mu )\frac{\partial _\mu}{\partial_\pm }X_{\sigma \pm }\partial _\sigma \,,
\eeq
\beq
\delta X_{\mu \pm }^2=2X_{\mu \pm }\partial _\pm \chi _\mu -X_{\sigma \pm }\,
\frac{\partial _\sigma }{\partial _\pm }\,2\,X_{\rho \pm }
(\partial _\pm \chi _\rho )-(\partial _\pm \chi _\sigma )
\frac{\partial _\sigma }{\partial _\pm }X_{\rho \pm }^2
\eeq
and
\beq
\delta \,\partial _\rho ...\partial _\rho ...=-(\partial _\pm  \chi _\mu )\,
\frac{\partial _\mu}{\partial _\pm }
\left(\partial _\rho ...\partial _\rho ...\right)\,.
\eeq
Here we neglected the terms canceled between various structure (see last contributions in
Eqs. (\ref{deltaX2}), (\ref{delXdel}) and (\ref{gmunu}))
\beq
\Delta \,\delta \left(X_{\sigma \pm} \partial _\sigma \right)=-X_ {\mu \pm}
 \partial _\mu  \chi _\sigma \partial _\sigma +\partial _\pm
 \chi _\mu \frac{\partial _\mu}{\partial_\pm }X_{\sigma \pm }\partial _\sigma \,,\,\,
 \Delta \,\delta X_{\mu \pm }^2=\partial _\pm \chi _\sigma
\frac{\partial _\sigma }{\partial _\pm }X_{\rho \pm }^2
\eeq
and
\beq
\Delta \,\delta \,\partial _\rho ...\partial _\rho ...=-\left(\partial _\rho \chi _\sigma
\partial _\sigma ...\right)
\partial _\rho -\partial _\sigma ...
\partial _\rho \chi _\rho
\partial _\rho +\partial _\pm  \chi _\mu \,\frac{\partial _\mu}{\partial _\pm }
\left(\partial _\rho ...\partial _\rho ...\right)\,.
\eeq
They generate unessential corrections to $j_{\pm \pm}$
proportional to $\partial _\pm$ 
\beq
\Delta \,j_{\pm }=-\partial _\pm \chi _\sigma
\partial _\sigma J_\pm \,.
\eeq

\section{Hamilton - Jacobi equation  for effective currents}

To construct covariant equations for the effective currents in all orders
we take into account,
that $j_{\pm \pm}$ are invariant under general coordinate transformations up to
total derivatives in $x^\pm$. Let us introduce
the currents $j^\mp$ related directly
to $j_{\pm \pm}$ 
\beq
j^\mp \equiv -\frac{1}{\partial _\pm}\,j_{\pm \pm}=J_\pm -\frac{1}{\partial _\pm}\,h_{\pm \pm} \,.
\label{jmp}
\eeq
Using these relations one can transform the "eikonal" equation (\ref{eqfan}) for $J_\pm ^{eik}$ to the form
\beq
-\left(\partial _\pm -h_{\sigma \pm } \partial _\sigma \right)j_{eik}^\mp =
h_{\pm \pm}-\left(h_{\rho \pm}\right)^2-
\frac{1}{4}\,\left(\partial _\rho j_{eik}^\mp \right)^2\,.
\eeq
In "eikonal" approximation the possible contributions containing the matrix
elements $h_{\mu \nu}$ with $\mu ,\nu \ne \pm$ are absent.
To restore such terms we should impose on the equation the property of general covariance.
To begin with, the inhomogeneous term can be modified in such a way, that it becomes proportional
to a matrix element of the contravariant metric tensor
\beq
h_{\pm \pm}-\left(h_{\rho \pm}\right)^2\rightarrow h_{\pm \pm}-g^{\rho \sigma }\,
h_{\rho \pm}\,h_{\sigma \pm}=
-g^{\mp \mp}\,.
\eeq
Here and later the tensors with covariant and contravariant indices are considered to be
different. They are related by a contraction with the metric tensor.

Using similar modifications for the linear and quadratic term, one can obtain the generally
covariant "fan" equation for the currents $j^\mp$
\beq
g^{\mp \sigma}\,\partial _\sigma j^\mp =g^{\mp \mp}+
\frac{g^{\rho \sigma}}{4}\left(\partial _\sigma  j^\mp \right)
\left(\partial _\rho  j^\mp \right)\,.
\label{faneqj}
\eeq

In an accordance with the general covariance the currents $j^\mp$ are
transformed as follows
\beq
\delta j ^\mp =2\,\chi ^\mp +\chi ^\rho \partial _\rho j^\mp \,,
\eeq
where the infinitesimal parameters $\chi ^\mp$ and $\chi ^\rho$ tend to zero at large $x$
in an accordance with the fact, that $j^\mp$ are defined
up to the contributions vanishing at $x^\pm \rightarrow \infty$. Indeed, the induced part of the
effective
action with an integration over $x^\pm$ can be written as follows
\beq
\Delta S=-\frac{1}{2\kappa }\, \left(\int d^2x_\perp dx^-\,
\lim _{x^+\rightarrow \infty }\left(j^-\,\frac{\partial ^2_\mu A^{++}}{2}\right)+
\int d^2x_\perp dx^+\,\lim _{x^- \rightarrow \infty}\left(j^+\,
\frac{\partial ^2_\mu A^{--}}{2}\right)\right)
\eeq
and this expression is not changed under such transformations.

The equation for $j^\mp$ can be presented in a simpler form
\beq
g^{\rho \sigma }\left(\frac{1}{2}\,\partial _\rho j^\mp -g_{\rho}^ \mp\right)\,\left(\frac{1}{2}\,
\partial _\sigma j^\mp -g_\sigma ^{\mp}\right)
=0\,.
\label{fanfull}
\eeq

Its formal solution  is
\beq
j^\mp=2\,x^\mp -\omega ^\pm \,,
\eeq
where the quantities $\omega ^\pm $ satisfy the light front equation
\beq
g^{\rho \sigma }\,\partial _\rho \omega ^\pm \,\partial _\sigma \omega ^\pm =0\,.
\eeq

The last equation can be obtained in an independent way if we would search the solution
of the generally covariant d'Alambert equation (see eq. (\ref{Dalamber}) in Section 4)
\beq
\nabla ^2\phi (x)=0
\eeq
for the wave function of the scalar particle moving with a large momentum $p^\mp$ in the semiclassical 
ansatz
\beq
\phi  ^\pm (x)=\exp \left(-i |p|\, x^\mp +i\,\theta ^\mp (x) \right)\,,\,\, \theta ^\mp (x)=\frac{|p| }{2}\,
j ^\mp (x)\,,
\eeq
where $\theta ^\mp (x)$ is a rapidly changing phase and $j ^\mp$ is the effective current
in our normalization. Indeed, by neglecting the derivatives from the metric tensors in comparison with
large derivatives from $\phi ^\pm$ we obtain from the d'Alambert  equation its semiclassical version
\beq
g^{\rho \sigma }\left(\frac{1}{2}\,\partial _\rho j^\mp -g_{\rho}^ \mp\right)\,\left(\frac{1}{2}\,
\partial _\sigma j^\mp -g_\sigma ^{\mp}\right)
=0\,,
\eeq
which coincides with the equation  (\ref{fanfull}) for $j^\mp$ derived above.

The $S$-matrix for the particle scattering at a given impact parameter off the gravitational field
in the semiclassical
approximation has the following form
\beq
S=\lim _{x^\pm \rightarrow \infty }\exp \left(i\frac{|p|}{2}\,j ^\mp (x)\right)\,,
\eeq
providing that the initial conditions for $j^\mp$ are
\beq
\lim _{x ^\pm \rightarrow -\infty} j ^\mp (x)=0\,.
\eeq
In particular this $S$-matrix contains pure eikonal contributions for which the particle in the intermediate
states lies on mass shell. Such contributions should be absent in the effective action, although they are reproduced
by the iteration of effective vertices in the $s$-channel. It is the reason, why the effective current
$j^\mp$ entering in the action at large $x^\pm$ is proportional to the logarithm of the $S$-matrix
\beq
\lim _{x^\pm \rightarrow \infty} \,j ^\mp (x)=-i\,\frac{2}{p^\pm}\,\ln S\,.
\eeq

It is well known~\cite{Fock}, that the solution of the light front equation
\beq
g^{\rho \sigma }\,\partial _\rho \omega  \,\partial _\sigma \omega =0
\eeq
can be expressed
in terms of the null-geodesic trajectories of particles in the gravitational field satisfying the equation
of motion
\beq
\frac{d^2x^\mu }{(d\tau )^2}=\Gamma ^\mu _{\alpha \beta} \,\frac{dx^\alpha }{d\tau}\,\frac{dx^\beta }{d\tau }
\,,
\eeq
where $\tau$ is a parameter increasing along the trajectory
and $\Gamma ^\mu _{\alpha \beta}$ is the Christoffel symbol. The geodesic equation is presented in the form of
the Hamilton equations
\beq
\frac{dx^\mu }{d\tau }=g^{\mu \nu}\,\omega _\nu \,,\,\,
\frac{d \omega _\alpha }{d\tau}=-\frac{1}{2}\,\omega _\mu \,\omega _\nu \partial _\alpha \,g^{\mu \nu}\,,
\eeq
where
\beq
\omega _\alpha \equiv \partial _\alpha \omega =g_{\alpha \beta }\frac{dx^\beta}{dp}
\eeq
plays a role of the particle momentum.

Note, that the light front equation can be considered as the Hamilton-Jacobi (HJ) equation for
the action $\omega$. Its general integral  contains an arbitrary function, but it is well known~\cite{Landau},
that this general solution is expressed in terms of the so-called complete integral containing only
4 arbitrary constants
\beq
\omega =a\,f(x^\mu ,c_1,c_2)+A\,.
\eeq
The appearance of the parameters $a$ and $A$ is related to the locality and the homogeneity of the HJ equation
(its invariance under the transformation $\omega \rightarrow b \,\omega +B$). Really the HJ equation is
an integral of motion for the Hamilton equations allowing to find the canonical variables
$x^\mu$ and $\omega _\nu$ as some functions of $\tau$. Indeed, providing that the HJ equation is fulfilled
at some  $\tau =\tau_0$ it will be valid at arbitrary $\tau$ due to the relation
\beq
\frac{d}{d\tau}\,g^{\rho \sigma }\,\partial _\rho \omega  \,\partial _\sigma \omega =0\,,
\eeq
which follows from the Hamilton equations.
On the contrary, the
general solution of the Hamilton equations can be obtained in terms of the complete integral for $\omega$.
Indeed,
one can prove~\cite{Landau}, that the derivatives of $\omega$ over the parameters $a,c_1,c_2$
\beq
\frac{\partial \omega}{\partial a}=f=d\,,\,\,\frac{\partial \omega}{\partial c_1}=d_1\,,\,\,
\frac{\partial \omega}{\partial c_2}=d_2
\eeq
are also integrals of motion and therefore one can find from the last relations the coordinates
$x^{i}$ ($i=1,2,3$) as functions of $t$ and six parameters $a,c_1,c_2,d,d_1,d_2$, which corresponds
to a general solution of equations of motion.

To construct a complete integral $\omega$ for our case of the massless particle scattering
off the gravitation field from the solution of the
Hamilton
equations we write the light front surface for arbitrary $\tau$ in the form
\beq
\omega (x^0,x^1,x^2,x^3)=const \,.
\eeq
Let us assume, that at large distances  and large negative times $t_0$, where $g^{\mu \nu}=\eta ^{\mu \nu}$,
this surface is a plane containing the points parametrised by two numbers $u$ and $v$
\beq
\vec{x}=t_0\vec{n}+u\,\vec{n}_1+v\,\vec{n}_2\,,
\eeq
where $\vec{n},\vec{n}_1,\vec{n}_2$ are orthogonal unit vectors
\beq
\vec{n}^2=\vec{n}^2_1=\vec{n}^2_2\,,\,\,(\vec{n},\vec{n}_1)=(\vec{n},\vec{n}_2)=(\vec{n}_1,\vec{n}_2)=0\,.
\eeq
The initial values of momenta are given below
\beq
\vec{\nabla}\omega =c\vec{n}\,,\,\,
\omega _0^2=c^2\,,
\eeq
where $c$ is an arbitrary parameter which depends generally on $u$ and $v$ (note, that $\omega$
is defined up to a common factor).
Then from the Hamilton equations one can calculate $x^\alpha$ and
$\omega _\beta$  for all values of $\tau$ and parameters $u$ and $v$.
Thus, we can obtain $x^\alpha =x^\alpha (\tau, u,v,t_0;\vec{n})$, which is a parametrised form
of the light front surface $\omega ^{(n)}=const$, depending on the light-cone vector
\beq
n=\frac{1}{\sqrt{2}}(1,\vec{n}).
\eeq
In the usual form this surface can be obtained by excluding the initial data $(u,v,t_0)$ from
four components of the vector $x^\alpha$.

In particular, to obtain the effective currents $j^{\pm}$ we should put
\beq
\omega ^{\pm}=\omega ^{(n^\pm )}\,,\,\,n^\pm =\frac{1}{\sqrt{2}}(1,\pm 1,0,0)
\eeq
and normalize the functions $\omega ^{\pm}$ in such a way, that
\beq
\omega ^\pm =2x^\pm -j^\pm \,.
\eeq
A possible generalization of the developed effective field theory could include a
superposition of the currents $j^{n}$ with different light-cone vectors $n$.

The classical equations for the effective actions apart from the usual Einstein term $G^{\sigma \rho}$
contain the induced terms for
the components of the metric tensor $g^{\mp \mp}$, $g^{\sigma \mp}$ and $g^{\sigma \rho}$. These terms
are equal to  the corresponding functional derivatives  of the action $\Delta S$. The
contributions proportional to $A^{++}$ and $A^{--}$ contain
the derivatives from the currents $j^{-}(x)$ and $j^+(x)$, respectively. Due to the Hamilton-Jacobi equations
these derivatives satisfy the relations
\beq
2g^{\rho  \sigma}\,(\partial _\rho \omega ^\mp )\,\partial _\sigma \frac{\delta j^\mp (x)}{\delta g^{\mu \nu} (y)}=
(\partial _\mu \omega ^\mp )\,(\partial _\nu \omega ^\mp )\,\delta ^4(x-y)\,.
\eeq
The induced terms in the Einstein equation play role of the energy-momentum tensor
$T_{\mu \nu} (y)$ which is conserved due to the general covariance of the action $\Delta S$.

\section{Effective action for shock wave gravitational fields}

To illustrate the general approach based on the effective action, let us consider  the Hamilton-Jacobi
equation for the massless particle scattering
off the gravitation center with the  metric tensor given by the Schwarzschild solution~\cite{Schwarz}
\beq
d^2s=\left(\frac{r-\alpha}{r+\alpha}\right)\,d^2t-\left(\frac{r+\alpha}{r-\alpha}\right)\,d^2r-
(r+\alpha )^2\left(d^2\theta +\sin ^2\theta \,d^2\phi\right)\,,
\eeq
where we used the spherical coordinates. The parameter  $\alpha$ is proportional to the mass $m$ of
the attraction center
\beq
\alpha =\gamma \,m\,,\,\,\kappa ^2=8\,\pi \,\gamma \,.
\eeq
The Einstein equations for the massless particle moving around the central body in the plane
$(x,y)$ corresponding to  $\theta =\pi /2$
are reduced to two ordinary differential equations~\cite{Fock}
\beq
\left(\frac{dr}{d\phi}\right)^2=\frac{(r+\alpha )^4}{b^2}-(r^2-\alpha ^2)\,,
\eeq
and
\beq
\frac{(r+\alpha )^3}{r-\alpha }\,\frac{dr}{dt}=\sqrt{(r+\alpha )^4-(r^2-\alpha ^2)\,b^2}\,,
\eeq
where $b$ in our case is the impact parameter of the colliding
particle which moves for $t\rightarrow -\infty$ along the line parallel to the axes $x$,
which corresponds
to the following initial condition for the first equation, describing its trajectory,
\beq
r(\phi )_{|_{\phi \rightarrow 0}}\approx \frac{b}{\phi} \rightarrow \infty\,.
\eeq
The solution of this equation can be expressed in terms of the elliptic integral of the first kind
\beq
\int _r^\infty \frac{b\,dr }{\sqrt{(r+\alpha )^4-(r^2-\alpha ^2)\,b^2}}=\phi \,.
\eeq
It allows to find $r$  as a function of $\phi$ and $b$. Inverting this function, one can find $b$
\beq
b=b(r,\phi ;\alpha)\,.
\eeq
The solution of the second equation can be written in the form
\beq
f(t,r,b,\alpha )\equiv t-\int ^r_0\frac{dr}{\sqrt{(r+\alpha )^4-(r^2-\alpha ^2)\,b^2}}
\,\frac{(r+\alpha )^3}{r-\alpha }=
C\,,
\eeq
where the constant $C$ is found from the initial conditions for $r$ at $t\rightarrow -\infty$.
In an accordance with our normalization we can construct the complete integral for the Hamilton-Jacobi
equation
\beq
\omega ^{(n)}=
2\,f(t,r,b(r,\phi ;\alpha ),\alpha )\,,
\eeq
where the unit vector $\overrightarrow{n}$ defines the direction of the initial particle
momentum and the impact parameter vector $\overrightarrow{\rho}$ is orthogonal to it. The angle $\phi$
is in fact the polar angle with the respect to the vector $\overrightarrow{n}$.

To obtain the effective currents $j^{\mp}$ we should put $\overrightarrow{n}=\mp \overrightarrow{e}_3$
and write $\omega$
in the form
\beq
\omega ^{\mp }=\omega ^{(n^\mp )}=2x^\mp -j^\mp \,.
\eeq

To simplify the perturbative expansion of the effective currents we consider below
the massless particle scattering off the gravitation center
moving with the relativistic velocity $v\rightarrow c$ in the direction of the third
axes $\overrightarrow{e}_3$.
Due to the Lorentz contraction the field of this center is given by the metric corresponding to
the shock wave solution of Aichelburg and Sexl
\beq
(ds)^2=\eta _{\mu \nu}\,dx^\mu\,dx^\nu +h_{--}(dx^- )^2 \,
\eeq
where
\beq
h_{--}=\frac{8}{\sqrt{2}}\,G\,\mu \ln |\overrightarrow{x}| \,\delta (x^-)\,,
\label{shock}
\eeq
where $\overrightarrow{x}$ is the transverse component of the vector $x^\rho$.

The Hamilton equation for the particle moving in this field has the form
\beq
\frac{dx^\mu }{d\tau }=\eta ^{\mu \nu}\omega _\nu -\delta ^\mu _+\,h_{--}\omega _+\,,\,\,
\frac{d \omega _\alpha }{d\tau}=\frac{1}{2}\,\omega _+ \,\omega _+\partial _\alpha \,h_{--}\,.
\eeq
Before reaching the shock wave the particle propagates along the
straight line
\beq
x^\mu =x^\mu _0+\omega _0^\mu \tau \,,\,\,\omega _0^\mu =\eta ^{\mu \nu}\,(\omega _\nu )_0 \,,\,\,
\eta _{\mu \nu}\omega _0^\mu \omega _0^\nu =0\,,
\eeq
where $x_0^\mu $ and $(\omega _\nu )_0$ are initial values of coordinates and momenta.
The collision with the moving plane $x^-=0$ takes place at the moment $\tau_c$ fixed by the
equation
\beq
z_c=t_c\,,
\eeq
where the coordinates of the particle are
\beq
z_c=z_0+\omega ^3_0\,\tau _c\,,\,\,t_c=t_0+\omega ^0_0\,\tau _c\,,\,\,
\overrightarrow{\rho } =\overrightarrow{x}_0-\overrightarrow{\omega }_0\,\tau _c\,.
\eeq
Here we introduced the notation $\overrightarrow{\rho} $ for the transverse coordinate
$\overrightarrow{x}$ at $\tau =\tau _c$.

At $\tau >\tau _c$ the new values of $\omega _\alpha $ are
\beq
\omega _+=(\omega _+)_0
\,,\,\,
\omega _-=(\omega _-)_0+\frac{4}{\sqrt{2}}\,\omega _+\,
G \,\mu \ln \rho \,\delta (x^-)
\,,\,\,
\overrightarrow{\omega}  =\overrightarrow{\omega} _0+\frac{4}{\sqrt{2}}\,
G\,\mu \frac{\overrightarrow{\rho}}{\rho ^2} \,\omega _+\,,
\eeq
where $\overrightarrow{\rho}$ is fixed by the initial conditions.

From the equation for $x^\mu$ we obtain
\[
\omega _\rho \,\omega _\sigma \,\frac{dg^{\rho \sigma}}{d\tau }=
\omega _\rho \,\omega _\sigma \,\left(
\eta ^{\mu \nu}\omega _\nu -\delta ^\mu _+\,h_{--}\omega _+\right)\,
\frac{d g^{\rho \sigma}}{dx^\mu }
\]
\beq
=-\omega _+^2\left(-\frac{8}{\sqrt{2}}
\,G\,\mu \frac{\overrightarrow{\omega}\overrightarrow{\rho}}{\rho ^2}\,\delta (x^-)+
\omega _+\,\frac{8}{\sqrt{2}}
\,G\,\mu \ln \rho \,\partial_-\delta (x^-)\right)\,.
\eeq
This relation is compatible with the Hamilton-Jacobi equation
\beq
g^{\rho \sigma}\,\omega _\rho \,\omega _\sigma =0\,,
\eeq
which can be verified by its differentiation in $\tau$ with the use of the
Hamilton equation for $\omega _\alpha$. From the
above explicit expressions for $\omega _\mu$ we derive also,
that the metric tensor $g^{\rho \sigma}$, calculated in the points
of the particle trajectory $x^\mu =x^\mu (\tau)$ is
\beq
g^{\rho \sigma}=\eta ^{\rho \sigma}-\delta ^\rho _+\delta ^\sigma _+ \frac{8}{\sqrt{2}}
\,G \,\mu \left( \ln \rho \,\delta (x^-)-
\left(\frac{\overrightarrow{\rho }\overrightarrow{\omega}_0}{\rho ^2\,\omega _+}
+\frac{2}{\sqrt{2}}\,
G\,\mu \frac{1}{\rho ^2}\,\theta (x^-)\right)\theta (x^-)\right) \,.
\eeq

The coordinates of the massless particle are
\beq
x^\pm =x^\pm _c+\omega ^\pm _0(\tau -\tau _c)\,,\,\,\overrightarrow{x}=\overrightarrow{\rho}
-\overrightarrow{\omega}_0(\tau -\tau _c)
\eeq
before its collision with the plane wave and
\[
x^-=\omega _+(\tau -\tau_c) \,,\,\,\overrightarrow{x}=
-\left(\frac{\overrightarrow{\omega} _0}{\omega_ +}+\frac{4}{\sqrt{2}}\,
G\,\mu \frac{\overrightarrow{\rho}}{\rho ^2} \right)\omega _+(\tau -\tau_c )+\overrightarrow{\rho}\,,
\]
\beq
x^+ =x^+_c+\left(\frac{\omega ^+_0}{\omega _+}
+\frac{4}{\sqrt{2}}\,G\,\mu \left(
\frac{\overrightarrow{\omega} _0\overrightarrow{\rho}}{\omega _+\rho ^2}+
\frac{2}{\sqrt{2}}\, G\,\mu \frac{1}{\rho ^2}\right)\right)\,\omega _+(\tau -\tau _c)-
\frac{4}{\sqrt{2}}\,G\,\mu\,\ln \rho
\eeq
after its collision at $\tau>\tau_c$. Here we used the identity
\beq
\int dx^-\,\theta (x^-)\,\delta (x^-)=\frac{1}{2}\,.
\eeq
Note, that the particle moves along the light ray
$g_{\rho \sigma}dx^\rho dx^\sigma =0$.

Finding $\tau -\tau _c$ and $\rho$
from two first equations and putting the result in the right hand side of
third relation we obtain the complete integral for the corresponding Hamilton-Jacobi equation
in our normalization
\beq
\omega =2x^+=2x^+_0+2\omega ^+_0(\tau -\tau _c)+j,
\eeq
where the effective current
\beq
j=\frac{8}{\sqrt{2}}\,G\,\mu\,\left(
\frac{\overrightarrow{\omega} _0\overrightarrow{\rho}}{\omega _+\rho ^2}\,x^-+
\frac{2}{\sqrt{2}}\, G\,\mu \frac{1}{\rho ^2}\,x^--\ln \rho \right)\,.
\eeq
Note, that this current can be written as follows
\beq
j=-2\,\frac{\omega ^+_0}{\omega _+}\,x^-+\frac{(\overrightarrow{\rho}-
\overrightarrow{x})^2}{x^-}-\frac{8}{\sqrt{2}}\,G\,\mu\,\ln \rho
\eeq
and the equation for $\overrightarrow{\rho}$ is simplified
\beq
\overrightarrow{\partial} \,j=2\,\frac{\overrightarrow{\omega}_0}{\omega _+}\,.
\label{eqrho}
\eeq

Let us consider the simplest case when the particle colliding with the shock wave
has the following initial conditions
\beq
\overrightarrow{\omega}_0=\omega ^+_0=0\,.
\eeq
In this case we have for the effective current
\beq
j^+=j(g)=\frac{8}{\sqrt{2}}\,G\,\mu\,\left(
\frac{2}{\sqrt{2}}\, G\,\mu \frac{x^-}{\rho ^2}-\ln \rho \right)\,.
\eeq
where the vector $\overrightarrow{\rho}$ satisfies the equation
\beq
\overrightarrow{x}=
\overrightarrow{\rho}\left(1-\frac{4}{\sqrt{2}}\,G\,\mu \,\frac{x^-}{\rho ^2}\right)\,.
\eeq
Its solution is given below
\beq
\overrightarrow{\rho}=\overrightarrow{x} \,f(z)\,,\,\,z=\frac{8}{\sqrt{2}}\,G\,\mu \,\frac{x^-}{|x|^2}\,,
\eeq
where
\beq
f(z)=\frac{1}{2}\,\left(1+\sqrt{1+2z}\right)=1+\frac{z}{2}-\frac{z^2}{4}+
\frac{z^3}{4}-\frac{5z^4}{16}+...\,.
\eeq

The current $j^+$ can be written in the form
\beq
j^+=a\,\left(\frac{1}{4}\,\frac{|x|^2}{|\rho|^2}\,z-\ln \rho \right)=
-a\,\left(\ln x+\phi (z)\right)\,,\,\,a=\frac{8}{\sqrt{2}}\,G\,\mu
\eeq
where
\beq
\phi (z)=\ln f(z)-\frac{1}{4}\,\frac{z}{f^2(z)}=\frac{z}{4}-\frac{z^2}{8}+\frac{5}{48}\,z^3
-\frac{7}{64}\,z^4+...\,\,.
\eeq
On the other hand, using expressions (\ref{eikcur1}), (\ref{eikcur2}) and (\ref{eikcur5}) for
the eikonal currents $j_{\pm \pm}^{eik}$ and (\ref{Xsigma}) for $X_{\sigma \pm}$ we can write
the current $j^+$ (\ref{jmp}) for the shock wave field (\ref{shock}) in the form
\[
j^+=-a\,\ln x +\frac{a^2}{\partial_-}\left(\frac{x_\sigma}{2x^2}\right)^2
-\frac{a^3}{\partial_-}\frac{x_\mu}{2x^2}\frac{\partial_\mu}{\partial_-}
\left(\frac{x_\sigma}{2x^2}\right)^2
+\frac{a^4}{\partial_-}
\frac{x_\nu}{2x^2}\frac{\partial_\nu}{\partial_-}
\frac{x_\mu}{2x^2}\frac{\partial_\mu}{\partial_-}
\left(\frac{x_\sigma}{2x^2}\right)^2
\]
\[
+\frac{a^4}{4\partial_-}\left(\frac{\partial_\mu}{\partial_-}
\left(\frac{x_\sigma}{2x^2}\right)^2\right)^2
-\frac{a^5}{\partial_-}\frac{x_\rho}{2x^2}\frac{\partial_\rho}{\partial_-}
\frac{x_\nu}{2x^2}\frac{\partial_\nu}{\partial_-}
\frac{x_\mu}{2x^2}\frac{\partial_\mu}{\partial_-}
\left(\frac{x_\sigma}{2x^2}\right)^2
\]
\beq
-\frac{a^5}{4\partial_-}\frac{x_\nu}{2x^2}\frac{\partial_\nu}{\partial_-}
\left(\frac{\partial_\mu}{\partial_-}
\left(\frac{x_\sigma}{2x^2}\right)^2\right)^2
-\frac{a^5}{2\partial_-}\left(\frac{\partial_\mu}{\partial_-}
\frac{x_\nu}{2x^2}
\frac{\partial_\nu}{\partial_-}
\left(\frac{x_\sigma}{2x^2}\right)^2
\right)\left(\frac{\partial_\mu}{\partial_-}
\left(\frac{x_\sigma}{2x^2}\right)^2\right)\,.
\eeq
Differentiating over $x_\sigma $ and integrating over $x^+$ we
obtain the same expression for $j^-$, which can serve as a
verification of the approach.

Let us consider now a more general situation of the massless scattering off
the gravitational field with the metric
\beq
g_{\mu \nu}=\eta _{\mu \nu}+\delta ^-_\mu \delta ^-_\nu V(\overrightarrow{x})\,\delta (x^-)\,,
\eeq
where the potential $V$ is an arbitrary function of the points
on the shock plane. Repeating the above calculation, we obtain the
generalized equation for the point $\overrightarrow{\rho}$ in which the particle crosses
the plane
\beq
\overrightarrow{x}=
\overrightarrow{\rho}-\frac{x^-}{2}\,\overrightarrow{\partial}\,V(\overrightarrow{\rho})
\eeq
and the expression for the effective current $j^+$
\beq
j^+=-V(\overrightarrow{\rho})+
\frac{x^-}{4}\,\left(\overrightarrow{\partial}V(\overrightarrow{\rho})\right)^2
=-V(\overrightarrow{\rho})+
\frac{(\overrightarrow{\rho}-\overrightarrow{x})^2}{x^-}\,.
\eeq
Note, that the equation for the point $\overrightarrow{\rho}$ can be written as the
stationarity condition for $j^+$ as a function of $\overrightarrow{\rho}$
\beq
\overrightarrow{\partial}\,j^+=0\,.
\eeq
Using the perturbation theory for the solution of the equation for
$\epsilon _\mu = \rho _\mu -x_\mu$ in metric $\eta _{\mu \nu}$
\beq
\epsilon _\mu =\frac{x^-}{2}V_\mu -\frac{x^-}{2}V_{\mu \mu _1}\frac{x^-}{2}V_{\mu_1}+
2\,\frac{x^-}{2}V_{\mu \mu _1}\frac{x^-}{2}V_{\mu _1\mu_2}\frac{x^-}{2}V_{\mu_2}+...\,,
\eeq
where
\beq
V_{\mu_1,\mu _2,...\mu_n}\equiv
\partial_{\mu _1}\partial_{\mu _2}...,\partial_{\mu _n}V(\overrightarrow{x})\,,
\eeq
and putting the result in $j^+$, we find
\beq
j^+_{eik}=-V(x)+ x^-\left(\frac{1}{2}V_\sigma \right)^2
+\frac{(x^-)^2}{2}V_\mu \partial _\mu \left(\frac{1}{2}V_\sigma \right)^2+...
\eeq
in an agreement with the expressions (\ref{eikcur2}) for the eikonal contribution
with the simplified expression for $X_{\sigma -}$
\beq
X_{\sigma -}\rightarrow -\frac{\partial_\rho}{\partial _-}\,g_{--}.
\eeq

\section{Variational principle for the effective currents}

Let us  consider even more general configuration of the gravitational field consisting from
$n$ shock waves moving in the $z$-direction
\beq
g_{\mu \nu}=\eta _{\mu \nu}+\delta ^-_\mu \delta ^-_\nu \,\sum _{r=1}^nV^{(r)}(\overrightarrow{x})
\,\delta (x^--x^-_r)\,,
\eeq
where $x^-_r$ are some parameters ordered in the following way
\beq
x_1^-<x_2^-<...<x_n^-\,.
\eeq
By solving the Hamilton equations for the massless particle flying at $\tau \rightarrow -\infty$ along the
$z$-axes from $z=-\infty$ with the impact parameter $\overrightarrow{\rho}$  and
$\omega ^+_0=\overrightarrow{\omega}_0=0$ for each of the intervals
$x^-_r<x^-<x^-_{r+1}$ for $r=1,2,...,n$ we obtain for the points $\overrightarrow{\rho}_r$
in which the trajectory crosses the corresponding planes the following recurrence relation
\[
\overrightarrow{\rho}_1=\overrightarrow{\rho}\,,\,\,\overrightarrow{\rho}_2=\overrightarrow{\rho}_1-
\frac{x^-_2-x^-_1}{2}\,\overrightarrow{\partial} _1\,V^{(1)}(\overrightarrow{\rho}_1)\,,\,\,
\overrightarrow{\rho}_3=\overrightarrow{\rho}_2-
\frac{x^-_3-x^-_2}{2}\,\sum _{t=1}^2\overrightarrow{\partial} _t\,V^{(t)}(\overrightarrow{\rho}_t)\,,...
\]
\beq
\overrightarrow{\rho}_n=\overrightarrow{\rho}_{n-1}-
\frac{x^-_n-x^-_{n-1}}{2}\,\sum _{t=1}^{n-1}\overrightarrow{\partial} _t\,V^{(t)}(\overrightarrow{\rho}_t)
\,,\,\,\overrightarrow{x}=\overrightarrow{\rho}_{n}-
\frac{x^--x^-_{n}}{2}\sum _{t=1}^{n}\overrightarrow{\partial} _t\,V^{(t)}(\overrightarrow{\rho}_t)\,,
\label{eqrho}
\eeq
where $\overrightarrow{x}$ and $x^-$ are coordinates of the particle after its interaction with
all shock waves. Note, that the $x^-$-coordinate of the particle and its momentum $\omega_+$ are not
changed during collisions
\beq
x^-=\omega _+\tau +x^-_0\,.
\eeq
But the momenta $\omega _+$ and $\overrightarrow{\omega}$ are different in each interval
$x^-_r<x^-<x^-_{r+1}$
\beq
\omega _-=\frac{\omega_+}{2}\,\sum _{r=1}^nV^{(r)}(\overrightarrow{\rho}_r)\,
\delta (x^--x^-_r)\,,\,\,
\overrightarrow{\omega}=\overrightarrow{\omega}_r=\frac{\omega_+}{2}
\sum_{t=1}^r\overrightarrow{\partial}_tV^{(t)}(\overrightarrow{\rho}_t)\,.
\eeq
The metric tensor, calculated on the particle trajectory in this interval, has the form
\beq
g^{\rho \sigma}=\eta ^{\rho \sigma}-\delta _+^\rho \delta _+^\sigma
\sum_{r=1}^n\left(V^{(r)}(\overrightarrow{\rho}_r)\delta (x^--x^-_r)
-\frac{\theta ^2(x^--x^-_r)}{4}\sum _{t=1}^r
\left(\overrightarrow{\partial}_tV^{(t)}(\overrightarrow{\rho}_t)
\right)^2\right)\,,
\eeq
compatible with the integral of motion
\beq
g^{\rho \sigma}\omega _\rho \omega_\sigma =2\omega _+\omega _--\overrightarrow{\omega}^2
+\omega _+^2g^{++}=0\,.
\eeq
Note, that the total derivative of $g^{++}$ in $\tau$ is in an agreement with the Hamilton
equations
\beq
\frac{dg^{++}}{d\tau}=-\omega _+\partial _-g^{++}+\frac{2\,\overrightarrow{\omega}}{\omega ^2_+}\,
\frac{d \overrightarrow{w}}{d\tau}\,.
\eeq
The coordinate $x^+$ is also changed after each collision and after all collisions we have
\beq
x^+=x^+_0+\frac{1}{2}\sum _{r=1}^n (x^-_{r+1}-x^-_r)\left(\sum _{t=1}^r
\frac{\overrightarrow{\partial_t}}{2}V^{(t)}(\overrightarrow{\rho}_t)\right)^2-\frac{1}{2}\,
\sum _{r=1}^nV^{(r)}(\overrightarrow{\rho}_r)\,,
\eeq
where it is implied, that $\overrightarrow{\rho}_{n+1}=\overrightarrow{x}$ and
$x^-_{n+1}=x^-$.
Thus, we obtain for the corresponding effective current in the above gravitational field
the following expression
\beq
j^+=\sum _{r=1}^n (x^-_{r+1}-x^-_r)\left(\sum _{t=1}^r
\frac{\overrightarrow{\partial_t}}{2}V^{(t)}(\overrightarrow{\rho}_t)\right)^2-
\sum _{r=1}^nV^{(r)}(\overrightarrow{\rho}_r)\,,
\eeq
where it is assumed, that the points $\overrightarrow{\rho}_r$ are expressed in terms of
$\overrightarrow{x}$ and $x_-$ with the use of equations (\ref{eqrho}). Due to these equations
the effective current can be written even in a simpler form
\beq
j^+=\sum _{r=1}^n \frac{(\overrightarrow{\rho}_{r+1}-\overrightarrow{\rho}_r)^2}{x^-_{r+1}-x^-_r}-
\sum _{r=1}^nV^{(r)}(\overrightarrow{\rho}_r)\,,\,\,\overrightarrow{\rho}_{n+1}=\overrightarrow{x}\,,
\,\,x^-_{n+1}=x^-\,.
\eeq
Such form of the effective current gives a possibility to write the equations for
$\overrightarrow{\rho}_r$ as its stationarity conditions
\beq
\overrightarrow{\partial}_r\,j^+=0\,.
\eeq
One can verify the perturbative expansion of this effective current by comparing it with the
general expressions (\ref{eikcur2}) for the eikonal contribution.

Let us consider
the continuous limit of the scattering problem, assuming that the number of shock waves is
infinite and the distance between them tends to zero. In this case the metric tensor on the
particle trajectory is
\beq
g^{\rho \sigma}(\overrightarrow{x},x^-)=\eta ^{\rho \sigma}-\delta _+^\rho \delta _+^\sigma
\left(g^{++}(\overrightarrow{\rho},x^-)+\frac{\overrightarrow{\omega}^2}{\omega _+^2}\right)\,,
\eeq
where
\beq
\overrightarrow{\omega}=\omega_+\,\partial_-\overrightarrow{\rho}
\eeq
and $\overrightarrow{\rho}$ is considered to be a function of $x_-$ and $\overrightarrow{x}$
calculated with the use of the equation of motion for the colliding particle.
The effective current
can be written in the integral form
\beq
j^+=\int _{-\infty}^{x^-}dy^-\left(g^{++}(y^-,\overrightarrow{\rho}(y^-))+
(\partial _{-}\overrightarrow{\rho})^2\right)\,,
\label{varifunct}
\eeq
where the variable $y^-$ enumerates the shock waves.

This functional can be considered as a classical action for the particle moving in the
gravitational field, which allows to formulate the variational principle for
the effective current $j^+$. Indeed, $j^+$ should be calculated on the particle
geodesic trajectory $\overrightarrow{\rho}(x^-,\overrightarrow{x})$. The trajectory
are found from the stationarity conditions for this functional which have the form
of the non-relativistic Newton equations
\beq
2\,\partial ^2 _- \overrightarrow{\rho}=
\overrightarrow{\partial}\,g^{++}\,.
\eeq
Note, that the "potential" $g^{++}$ depends explicitly on $x_-$ and therefore the energy,
which is a formal integral of motion for this equation,
is not conserved. But with taking into account, that the partial derivatives in $x^-$ of
$\omega _-$ and $g^{++}$ are proportional,
we can write the correct integral of motion in the form
\beq
\left(\partial _-\overrightarrow{\rho}\right)^2-g^{++}-2\,\frac{\omega _-}{\omega _+}=0\,,
\label{motionint}
\eeq
which is really coincides with the Hamilton-Jacobi equation. Indeed,
the variation over $\overrightarrow{\rho}$ in the integrand for $j^-$ after the use of the
stationarity equations gives a total
derivative over $x^-$ leading after its integration to the relation
\beq
\delta j^+=2\,(\partial _-\overrightarrow{\rho})\,\delta  \overrightarrow{\rho}
\eeq
and therefore we have the relation
\beq
\partial _-\overrightarrow{\rho}=\frac{1}{2}\,\overrightarrow{\partial}\,j^+=
-\frac{\overrightarrow{w}}{\omega _+}\, .
\eeq
As a result, the integral of motion (\ref{motionint}) coincides with the HY equation
for this case.

Therefore we obtain the non-linear equation for $j^+$ compatible with the
above variational principle
\beq
j^+=\int _{-\infty}^{x^-}dy^-\left(g^{++}(y^-,\overrightarrow{\rho}(y^-))+
\frac{1}{4}\,\left(\overrightarrow{\partial}\,j^+\right)^2\right)\,.
\eeq
Here the functions $\overrightarrow{\rho}(y^-))$ are solutions of the Hamilton equations.
The iteration of this equation over $g^{++}$ reproduces results (\ref{eikcur2}) for
the pure eikonal contribution
\beq
j^+_{eik}=\frac{1}{\partial _-}\,g^{++}-\frac{1}{\partial _-}
\left(\frac{1}{2}\frac{\overrightarrow{\partial}}{\partial_-}
\,g^{++}\right)^2
+\frac{1}{\partial _-}\,\left(\frac{1}{2}\,\frac{\overrightarrow{\partial}}{\partial_-}
\,g^{++}\right)\,\frac{\overrightarrow{\partial}}{\partial _-}\,
\left(\frac{1}{2}\,\frac{\overrightarrow{ \partial}}{\partial_-}
\,g^{++}\right)^2+...\,,
\eeq
where the contributions from the expansion of $g^{++}$ in $\overrightarrow{\rho}$ with the subsequent
use of the hamilton equation and the integration by parts are also
taken into account.

As we argued in the previous sections, the effective currents $j^\pm$ as functionals
of the metric tensors in a general form satisfy the Hamilton--Jacobi equation (see (\ref{faneqj}))
\beq
g^{\sigma \pm}\partial_\sigma j^\pm =g^{\pm \pm}+\frac{g^{\sigma \rho}}{4}\,
\left(\partial _\sigma j^\pm \right)\,\left(\partial _\rho j^\pm\right)\,.
\eeq
It would be important to write the solution of this equation as an extremum of an
local functional similar to (\ref{varifunct}), because in the perturbative expansion
(\ref{effpert1}, \ref{effpert2}) this locality property is lost. Moreover, such functional could
help us in finding quantum-mechanical corrections to the effective action and its supersymmetric
generalization.
For this purpose one should present (\ref{varifunct}) in the form invariant under the general
covariant transformations. We hope to return to this problem in our future publications.

\section{Effective reggeon-graviton vertices}

Let us apply the effective action to the problem of calculations of the simplest effective vertices for the
reggeon-graviton interactions in the lowest order of the perturbation theory. For this purpose it is enough to
leave in the currents $j_{\pm \pm}$ only two first terms of the perturbative expansion
\beq
j_{\pm \pm} \approx h_{\pm \pm}-X_{\sigma \pm}^2\,,\,\,X_{\sigma \pm}=h_{\sigma \pm}-\frac{1}{2}\,
\frac{\partial _\sigma}{\partial _\pm}\,h_{\pm \pm}\,.
\eeq
We expand also the Christoffel symbol
\beq
\Gamma ^\rho _{\mu \nu}\approx \frac{1}{2}\left(\partial _\mu h_{\rho \nu}+\partial _\nu h_{\rho \mu}-
\partial _\rho h_{\mu \nu}-h_{\rho \sigma} (\partial _\mu h_{\sigma \nu}+\partial _\nu h_{\sigma \mu}-
\partial _\sigma h_{\mu \nu})
\right)
\eeq
and the Hilbert-Einstein Lagrangian
\beq
\sqrt{-g}\,R=\sqrt{-g}\,g^{\mu \nu} \left(\partial _\nu \Gamma ^\rho _{\mu \rho}-
\partial _\rho \Gamma ^\rho _{\mu \nu}+\Gamma ^\sigma _{\mu \rho}\Gamma^\rho _{\sigma \nu} -
\Gamma ^\sigma _{\mu \nu}\Gamma ^\rho _{\sigma \rho}\right)\approx L_2+L_3\,,
\eeq
where
\beq
L_2=\frac{\partial _\sigma h_{\mu \sigma }}{2}\left(\partial _\mu h_{\rho  \rho }-\partial _\rho h_{\mu \rho }\right)+
\frac{1}{4}\left((\partial _\sigma h_{\mu \nu })^2-(\partial _\sigma h_{\mu \mu })^2\right)
\eeq
and
\[
L_3=h_{\rho \sigma}\left((\partial_\mu h_{\mu \sigma})\partial_\nu h_{\nu \rho}-
\frac{\partial_\rho h_{\mu \nu}}{4}\,\partial_\sigma h_{\mu \nu}
-\frac{\partial_\mu h_{\nu \sigma}}{2}(\partial_\nu h_{\mu \rho}+\partial_\mu h_{\rho \nu})
+\frac{\partial _\mu h_{\mu \nu}}{2}\,
(2\partial_\rho h_{\nu \sigma}-\partial _\nu
h_{\rho \sigma })
\right)
\]
\[
+h_{\rho \rho}\,\left(h_{\mu \nu }(\partial _\mu \partial _\sigma h_{\nu \sigma}-\frac{1}{2}\,
\partial _\sigma ^2h_{\mu \nu})+\frac{1}{2}\,(\partial _\nu h_{\nu \sigma})^2
-\frac{3}{8}\,(\partial_\sigma h_{\mu \nu})^2 +\frac{1}{4}\,(\partial _\sigma h_{\mu \nu})\,
\partial_\mu h_{\sigma \nu}\right)
\]
\beq
-\frac{h_{\rho \rho}}{8}\left(\partial _\nu h_{\sigma \sigma}\right)^2
-\frac{h_{\rho \rho}}{4}\,h_{\mu \nu}\,\partial _\mu \partial_\nu h_{\sigma \sigma}\,.
\eeq
These expressions are valid up to the terms proportional to total derivatives which give
vanishing contributions to the action $S_{HE}$.

The action is invariant under the general coordinate transformations
\beq
\delta S_{HE}=0
\eeq
with the same accuracy,
which can be verified by checking the following relations
\beq
\delta L_2 = \partial _\nu \left(\frac{h_{\mu \mu}}{2}\,
(\partial_\sigma^2\chi_\nu-\partial_\nu\partial_\sigma \chi_\sigma )+h_{\rho \mu}
\partial_\rho \partial_\nu \chi_\mu -h_{\nu \mu}\partial_\rho^2\chi_\mu \right)+\chi_\mu
\,a_\mu\,,\,\,\delta L_3=-\chi_\mu \,a_\mu\,,
\label{invarL}
\eeq
where
\[
a_\mu \approx \frac{\partial_\sigma ^2h_{\nu \nu}}{2}\partial_\mu h_{\rho \rho}
-\left(\partial_\sigma ^2h_{\nu \nu}\right)\partial_\delta h_{\delta \mu}-
\frac{\partial_\mu h_{\nu \nu}}{2}\partial _\sigma \partial _\rho h_{\rho \sigma}
+\frac{\partial _\delta \partial_\nu h_{\rho \rho}}{2}\left(
\partial _\nu h_{\delta \mu}+\partial _\delta h_{\nu \mu}-\partial _\mu
h_{\delta \nu}\right)
\]
\beq
+(\partial _\sigma \partial_\nu h_{\sigma \nu})\,\partial_{\rho}h_{\rho \mu}
+\left(\frac{\partial_\sigma ^2h_{\rho \nu}}{2}-\partial _\rho \partial_\delta h_{\nu \delta}\right)\left(
\partial _\nu h_{\rho \mu}+\partial _\rho h_{\nu \mu}-\partial _\mu
h_{\rho \nu}\right).
\label{invarL23}
\eeq

Because the induced contributions to the action are also generally covariant,
the Euler-Lagrange equations for the total action are
self-consistent. We can write them in the form
\beq
R^{\mu \nu}-\frac{1}{2}\,g^{\mu \nu}\,R=\frac{1}{2}\,\frac{\delta}{\delta g_{\mu \nu}}
\int d^4x\, \left(j_{++}\partial_\sigma ^2A^{++}+j_{--}\partial_\sigma ^2A^{--}\right)
\,,
\eeq
where in the right hand side it is implied as usual, that the calculation of the variational
derivative over $g_{\mu \nu }$ is combined with the corresponding integration by parts.
The solution of these equations can be expanded in the series over the reggeon fields  $A_{\mp \mp} $
\beq
\overline{h}_{\mp \mp} =A_{\mp \mp} +O(A^2)
\eeq
similar to the case of the Euler-Lagrange equation for the effective action in QCD~\cite{eff}. Inserting
this solution in the effective action one can obtain various effective vertices for the self-interaction of the
reggeon fields $A_{\mp \mp}$ in the tree approximation. The physical gravitational fields will correspond to the
fluctuations $\delta h=h-\overline{h}$ around the classical solution. The functional integration over these fluctuations in
the quadratic approximation will lead to the
graviton Regge trajectories and to various reggeon couplings in the one-loop approximation.
This traditional approach will be considered in future publications. Here we restrict ourselves to the simple cases
where the results can be obtained in the lowest orders of perturbation theory.

To begin with, we note, that performing the functional gaussian integration over $h_{\mu \nu}$ from the exponent containing
the induced
action with the terms linear $h_{+ +}$ and $h_{--}$ we obtain the kinetic term
for the fields $A_{\pm \pm}$
\beq
-\frac{1}{2\kappa}\int \frac{d^4x}{2}\left(-\partial _\sigma h_{++}\,\partial _\sigma h_{--}-h_{++}\partial_\sigma ^2A_{--}
-h_{--}\partial_\sigma ^2A_{++}
\right)\rightarrow -\frac{1}{2\kappa}\int
d^4x\,\frac{\partial_\sigma A_{++}\,\partial _\sigma A _{--}}{2}\,.
\eeq
The kinetic term for the reggeon fields should have an opposite sign. Therefore we include in the
effective action the bare kinetic term for the reggeon fields
\beq
S_{kin}=\frac{1}{2\kappa}\int d^4x\,\partial_\sigma A_{++}\,\partial _\sigma A_{--}
\eeq
to have the correct renormalized contribution. Strictly speaking the propagator of the reggeized graviton should contain
the $\theta$-function corresponding to the ordering of rapidities $y$ in the different clusters
\beq
<0|\left(A^{y_1}_{--}(x_1)A^{y_2}_{++}(x_2)\right)=4\frac{\kappa}{\pi ^2} \,\theta (y_1-y_2)\,\frac{i}{(x_1-x_2)^2}\,.
\eeq

Further, the next order corrections in each of the induced actions
\[
-\frac{1}{2\kappa}\int \frac{d^4x}{2}\left(
-\left(h_{++}-\frac{1}{4} \left(\frac{\partial_\rho}{\partial _+}\,h_{++}\right)^2\right)\partial_\sigma ^2A_{--}
-\left(h_{--}-\frac{1}{4} \left(\frac{\partial_\rho}{\partial _-}h_{--}\right)^2\right)\partial_\sigma ^2A_{++}\right)
\]
lead with the use of the gaussian integration over the fields $h_{\pm \pm}$
to the cubic interactions of the reggeon fields
\beq
S^{1\rightarrow 2}= -\frac{1}{2\kappa}\int \frac{d^4x}{8}\, \left(\left(\frac{\partial _\rho}{\partial _+}
A_{++}\right)^2\,
\partial_\sigma ^2A_{--}+\left(\frac{\partial _\rho}{\partial _-}A_{--}\right)^2\,
\partial_\sigma ^2A_{++}\right)\,.
\eeq
Note, that the usual triple graviton vertex gives a vanishing contribution to this interaction.

In an analogous way one can calculate in the tree approximation the effective action for the reggeon
transitions $1\rightarrow n$
\beq
S^{1\rightarrow n}= -\frac{1}{2\kappa}\int \frac{d^4x}{2}\, \left(\partial _+J_+^{eik} (A_{++})\,
\partial_\sigma ^2A_{--}+\partial _-J_-^{eik} (A_{--})\,
\partial_\sigma ^2A_{++}\right)\,,
\eeq
where the "eikonal" currents $J_{\pm }^{eik}(h_{\pm \pm})$ can be obtained from the solution of
the "fan" equations
\beq
\left(\partial _\pm +\frac{1}{2}\,\left(\frac{\partial _\sigma  }{\partial _\pm }\,h_{\pm \pm}\right)
\,\partial_\sigma \right) J_{\pm}^{eik}=
\frac{1}{4}\,\left(\frac{\partial_\sigma}{\partial _\pm}\,h_{\pm \pm}\right)^{2}+
\frac{1}{4}\,\left(\partial _\rho J_\pm ^{eik }\right)^2\,.
\eeq

The effective action for the reggeon transitions $2\rightarrow n$ ($n\ge 2$) in the same
approximation contains a contribution from the usual triple graviton vertex. The general reggeon
interaction $n\rightarrow m$ is expressed in terms of
the solution of the Euler-Lagrange equation for the effective theory.

Let us consider now the effective action for the reggeon-reggeon-graviton (RRG) interaction in a tree
approximation $ S^{RRG}$. It contains the
contribution from the triple graviton vertex (gv) and from the second order (so) correction
($\sim h^2$) to the induced action
\beq
S^{RRG}=\frac{1}{2\kappa}\,\int d^4x L^{RRG}\,,\,\,
L^{RRG}= L^{RRG}_{gv} + L^{RRG}_{so} \,,
\eeq
where
\[
L^{RRG}_{gv}=A_{++}\left(
\left(\partial _\mu h_{\mu +}-\frac{\partial_- h_{++}}{2}\right)\,\partial_- A_{--}
-\left(\partial_+ h_{\mu -}+\partial _\mu h_{-+}+
\frac{\partial _\nu h_{\nu \mu}}{2} \right)\partial_\mu A_{- -}
\right)
\]
\[
+A_{--}\left(
\left(\partial _\mu h_{\mu -}-\frac{\partial_+ h_{--}}{2}\right)\,\partial_+ A_{++}
-\left(\partial_- h_{\mu +}+\partial _\mu h_{-+}+
\frac{\partial _\nu h_{\nu \mu}}{2} \right)\partial_\mu A_{+ +}
\right)
\]
\[
-h_{\rho \sigma}
\frac{\partial_\rho A_{++}}{2}\,\partial_\sigma A_{--}
-h_{+-}\,(\partial _-A_{--})\partial _+A_{++}-h_{+-} (\partial_\nu A_{+ +})\partial_\nu A_{--}
\]
\beq
+h_{\rho \rho}\,\left(-\frac{1}{2}\,A_{++}\,\partial_\sigma ^2A_{--}-\frac{1}{2}\,A_{--}\,\partial_\sigma ^2A_{++}
-\frac{3}{4}(\partial _\sigma A_{++})\partial _\sigma A_{--} +\frac{1}{2}(\partial _+A_{++})\partial _-A_{--}\right)\,.
\eeq
and
\[
L^{RRG}_{so}=\left(-\left(h_{- +}-\frac{1}{2}\,\frac{\partial _-}{\partial _+}h_{++}\right)A_{++}+
\frac{1}{2}\,\left(h_{\sigma +}-\frac{1}{2}\,\frac{\partial _\sigma}{\partial
_+}h_{++}\right)\,
\left(\frac{\partial _\sigma}{\partial _+}A_{++}\right)\right)\partial _\rho ^2A_{--}
\]
\beq
+\left(-\left(h_{- +}-\frac{1}{2}\,\frac{\partial _+}{\partial _-}h_{--}\right)A_{--}+
\frac{1}{2}\,\left(h_{\sigma -}-\frac{1}{2}\,\frac{\partial _\sigma}{\partial
_-}h_{--}\right)\,
\left(\frac{\partial _\sigma}{\partial _-}A_{--}\right)\right)\partial _\rho ^2A_{++}
\eeq

The effective action $S^{RRG}$ is invariant
\beq
\delta S^{RRG}=0
\eeq
under the "abelian" part of the general covariant transformation
\beq
\delta h_{\rho \sigma}=\partial _\rho \chi _\sigma +\partial _\sigma \chi _\rho
\eeq
because the corresponding contributions $S^{RRG}_{gv}$ and $S^{RRG}_{so}$ are transformed
as follows (cf. (\ref{invarL}) and (\ref{invarL23}))
\[
\delta S^{RRG}_{gv}=-\delta S^{RRG}_{so}=\frac{1}{2\kappa}\int d^4x \,\Phi (x)\,,
\]
\beq
\Phi (x)=\left(-\chi _-\partial_+A_{++}+\frac{1}{2}\,\chi _\sigma
\partial _\sigma A_{++}\right)\partial _\rho ^2A_{--}
+\left(-\chi _+\partial_-A_{--}+\frac{1}{2}\,\chi _\sigma \partial _\sigma A_{--}\right)
\partial _\rho ^2A_{++}\,.
\eeq
For the field of the produced graviton on the mass shell we have additional constraints
\beq
\partial _\mu ^2h_{\rho \sigma}=\partial _\mu h_{\mu \rho}=h_{\rho \rho}=0
\label{shellcond}
\eeq
and the RRG lagrangian is simplified as follows
\[
L^{RRG}=A_{++}\left(
-\frac{\partial_- h_{++}}{2}\,\partial_- A_{--}
-\left(\partial_+ h_{\mu -}+\partial _\mu h_{-+}\right)\partial_\mu A_{- -}
\right)
\]
\[
+A_{--}\left(
-\frac{\partial_+ h_{--}}{2}\,\partial_+ A_{++}
-\left(\partial_- h_{\mu +}+\partial _\mu h_{-+}\right)\partial_\mu A_{+ +}
\right)
\]
\[
-h_{\rho \sigma}
\frac{\partial_\rho A_{++}}{2}\,\partial_\sigma A_{--}
-h_{+-}\,(\partial _-A_{--})\partial _+A_{++}-h_{+-} (\partial_\nu A_{+ +})\partial_\nu A_{--}
\]
\[
+\left(-\left(h_{- +}-\frac{1}{2}\,\frac{\partial _-}{\partial _+}h_{++}\right)A_{++}+
\frac{1}{2}\,\left(h_{\sigma +}-\frac{1}{2}\,\frac{\partial _\sigma}{\partial
_+}h_{++}\right)\,
\left(\frac{\partial _\sigma}{\partial _+}A_{++}\right)\right)\partial _\rho ^2A_{--}
\]
\beq
+\left(-\left(h_{- +}-\frac{1}{2}\,\frac{\partial _+}{\partial _-}h_{--}\right)A_{--}+
\frac{1}{2}\,\left(h_{\sigma -}-\frac{1}{2}\,\frac{\partial _\sigma}{\partial
_-}h_{--}\right)\,
\left(\frac{\partial _\sigma}{\partial _-}A_{--}\right)\right)\partial _\rho ^2A_{++}
\eeq

Moreover, the corresponding RRG vertex can be written in
the momentum space as follows~\cite{Lip1, Lip2}
\beq
\Gamma _{\mu \nu}^{RRG}(q_2,q_1)=\frac{1}{2}\,C _\mu (q_2,q_1)\,C _\nu (q_2,q_1)
-\frac{1}{2}\,N_{\mu}(q_2,q_1)\,N_\nu (q_2,q_1)\,.
\eeq
Here $C (q_2,q_1)$ is the effective vertex describing the gluon
production from the reggeized gluon
\beq
C (q_2,q_1)=-q_1^\perp -q_2^\perp +p_A\left(\frac{q_1^2}{kp_A}+\frac{kp_B}{p_Ap_B}\right)-
p_B\left(\frac{q_2^2}{kp_B}+\frac{kp_A}{p_Ap_B}\right)\,,
\eeq
where $q_1, q_2$ are the momenta of the reggeized gluons, $k=q_1-q_2$
is the momentum of the produced gluon and $p_A,p_B$ are the momenta of the
colliding particles. The vector $N (q_2,q_3)$ is proportional to the photon bremstrahlung
factor in QED
\beq
N (q_2,q_1)=\sqrt{q_1^2q_2^2}\,\left(\frac{p_A}{p_Ak}-\frac{p_B}{p_Bk}\right)\,.
\eeq
Using the light-cone gauge for the polarization tensor of the produced graviton the RRG vertex
can be written in a simple form, which allows one to construct the corresponding term
in the effective action for the scattering amplitude with the multi-regge unitarity~\cite{Lip3}.

Let us consider now the effective action for the graviton scattering off the reggeized
gravitons. It can be written as a sum of two terms
\beq
S^{GGR}=\frac{1}{2\kappa} \int d^4x \left(L^{GGR}(A_{++})+L^{GGR}(A_{--})\right)\,,
\eeq
proportional to $A_{++}$ and $A_{--}$, respectively. We consider only the first term, because
the second one can be obtained from it by interchanging the light-cone indices $+$ and $-$.
In turn, $L^{GGR}(A_{++})$ is the sum of contributions from the triple reggeon vertex (rv) and
the second order (so) correction to the induced term
\beq
L^{GGR}(A_{++})=L^{A_{++}}_{gv}+L^{A_{--}}_{so}\,,
\eeq
where
\[
L^{A_{++}}_{gv}=
A_{++}\left((\partial_\mu h_{\mu -})^2-
\frac{(\partial_- h_{\mu \nu})^2}{4}
-\frac{\partial_\mu h_{\nu -}}{2}(\partial_\nu h_{\mu -}+\partial_\mu h_{\nu -})
+\frac{\partial _\mu h_{\mu \nu}}{2}\,
(2\partial_- h_{\nu -}-\partial _\nu
h_{-- })\right)
\]
\[
-(\partial _\rho A_{+ +})\left(\frac{h_{\rho \sigma} }{2}\, \partial _\sigma h_{--}+h_{\sigma -}
(\partial _-h_{\rho \sigma}+\partial_\rho h_{\sigma -})-h_{\rho -}\partial _{\sigma}h_{\sigma -}
+\frac{h_{--}}{2}\,\partial_\sigma h_{\sigma \rho}\right)
\]
\beq
+h_{\rho \rho}\,\left(-A_{++}\left(\frac{\partial _\sigma ^2h_{--}}{2}+
\frac{\partial _-^2h_{\sigma \sigma}}{4}\right)
-\frac{h_{--}}{2}\partial _\sigma ^2A_{++}
 +
(\partial _\sigma A_{++})\left(\frac
{\partial_- h_{\sigma -}}{2}-3\,\frac{\partial_\sigma h_{--}}{4}\right) \right)
\eeq
and
\beq
L^{A_{++}}_{so}=-
\frac{1}{2}\,\left(h_{\rho -}-\frac{1}{2}\frac{\partial _\rho}{\partial_-}h_{--}\right)^2\partial _\sigma ^2A_{++}\,.
\eeq
The corresponding lagrangians are transformed under the general coordinate transformations as follows
\[
\delta L^{A_{++}}_{gv}=-\chi_\rho \left(
(\partial_-h_{\rho -}-\frac{1}{2}\,\partial _\rho h_{--})\,\partial _\sigma ^2A_{++}
-(\partial _-^2h_{\sigma \sigma}+\partial_\sigma ^2h_{--}-2\partial_-\partial_\sigma h_{\sigma -})
\,\frac{\partial _\rho A_{++}}{2}\right)
\]
\beq
-\chi _-\,\left(\partial _-\partial _\rho h_{\sigma \sigma}+\partial _\sigma ^2-\partial _-\partial_\sigma
h_{\rho \sigma}-\partial_\rho \partial _\sigma h_{\sigma -}\right)\,\partial_\rho A_{++}
\eeq
and
\beq
\delta L^{A_{++}}_{so}=\chi _\rho \left(\partial _- h_{\rho -}-\frac{1}{2}\,\partial _\rho h_{--}\right)
\partial _\sigma ^2 A_{++}
\eeq
We can simplify the GGR lagrangian providing that gravitons are on the mass shell and their fields
satisfy additional constraints (\ref{shellcond})
\[
L^{GGR}(A_{++})=
A_{++}\left(-
\frac{(\partial_- h_{\mu \nu})^2}{4}
-\frac{\partial_\mu h_{\nu -}}{2}(\partial_\nu h_{\mu -}+\partial_\mu h_{\nu -})
\right)
\]
\beq
-(\partial _\rho A_{+ +})\left(\frac{h_{\rho \sigma} }{2}\, \partial _\sigma h_{--}+h_{\sigma -}
(\partial _-h_{\rho \sigma}+\partial_\rho h_{\sigma -})\right)
-
\frac{1}{2}\,\left(h_{\rho -}-\frac{1}{2}\frac{\partial _\rho}{\partial_-}h_{--}\right)^2\partial _\sigma ^2A_{++}\,.
\eeq
The corresponding vertex for the graviton scattering off the reggeon field $A_{++}$
can be written as follows (see ref.~\cite{Lip1, Lip2})
\beq
\Gamma  ^{GGR}_{\mu \nu ,\mu ' \nu'}=\frac{1}{2}\,\left(\Gamma _{\mu \mu '}^{GGR}\Gamma _{\nu \nu'}^{GGR}+
\Gamma ^{GGR}_{\mu \nu '}\Gamma ^{GGR}_{\nu \mu'}\right)\,,
\eeq
where $\Gamma _{\mu \mu '}^{GGR}$ is the effective vertex for the gluon scattering off the
reggeized gluon field $A_{+}$
\beq
\Gamma _{\mu \mu '}^{GGR}=-\left(\eta _{\mu \mu'}-
\frac{k'_\mu p^B_{\mu '}+k_{\mu'} p^B_{\mu }}{kp^B}-q^2\,\frac{p^B_{\mu }p^B_{\mu '}}{2(kp^B)^2}\right)\,,
\eeq
where $k$ and $k'$ are momenta of the initial and final gluons, $p^B$ is the momentum
of the another initial gluon and $q$ is the momentum transfer. After the transition to the helicity
basis the above vertex $\Gamma  ^{GGR}_{\mu \nu ,\mu ' \nu'}$ corresponds to the
conservation of the graviton helicity and leads to the corresponding contribution in the
effective action for the scattering amplitude with the multi-Regge unitarity~\cite{Lip3}.

\section{Graviton Regge trajectory and supergravity}

To calculate the graviton Regge trajectory in one loop~\cite{Lip1} it is needed to contract two GGR vertices
appearing in $L^{GGR}(A_{++})$ and $L^{GGR}(A_{--})$  with two
graviton propagators and integrate the product over the loop momentum. The integration over the
Sudakov variables $\alpha$ and $\beta$  of the virtual graviton momentum should give $\ln s$ equal to the
relative rapidity of the initial particles. To obtain a non-trivial $s$-dependence in each of two GGR lagrangians one should leave
only the singular contributions appearing in the induced terms
\beq
L^{GGR}(A_{\pm \pm}) \approx \left(\frac{1}{2}\,h_{\rho \mp}\,\frac{\partial _\rho}{\partial _\mp}h_{\mp \mp}-
\frac{1}{8}\,\left(\frac{\partial _\rho}{\partial _\mp}h_{\mp \mp}\right)^2\right)\,\partial_\sigma ^2A_{\pm \pm}\,.
\eeq
From these expressions one can derive the scattering amplitude described by the
contribution of the box diagrams corresponding to two
graviton exchange in the crossing channel
\beq
F=\delta _{\lambda _A \lambda _{A'}}\delta _{\lambda _B \lambda _{B'}}\,
\frac{\kappa ^4_L\,s^2}{(2\pi )^4 i} \int \frac{d^2k_\perp dk_+dk_-}{(k^2_\perp+2k_+k_-+i\epsilon )
((q-k)^2_\perp+2k_+k_-+i\epsilon )}\,f(k,q)\,,
\eeq
where $\lambda _r$ are the helicities of the scattered particles and the function $f(k,q)$ is given below
\[
f(k,q)=\frac{1}{2}\,\frac{(k ,q-k )^2}{(k_+k_-)^2}+\frac{k^2+(q-k)^2+4(k,q-k)}{k_+k_-}
\]
\beq
=\frac{(k_\perp ,q_\perp-k_\perp )^2}{4(k_+-i\epsilon )^2}
\left(\frac{1}{(k_-+i\epsilon )^2}+\frac{1}{(k_--i\epsilon )^2} \right)+\frac{q_\perp ^2}{2(k_+-i\epsilon )}
\left(\frac{1}{k_-+i\epsilon} +\frac{1}{k_--i\epsilon } \right)\,.
\eeq
Here we restored the analytic structure of the poles in an accordance with the Feynman $i\epsilon$-prescription.
The integral over $k_+$ in $F$ is non-zero only for $k_->0$. Taking it by residues with the
subsequent integration over $k_-$ one can obtain
\beq
F=F_{Born}\,\omega (t)\,\ln s\,,\,\,t=q_\perp ^2\,,
\eeq
where
\beq
F_{Born}=\delta _{\lambda _A \lambda _{A'}}\delta _{\lambda _B \lambda _{B'}}\,\kappa ^2\,\frac{s^2}{t}
\eeq
is the scattering amplitude in the Born approximation
and $j=2+\omega (t)$ is the graviton Regge
trajectory~\cite{Lip1}
\beq
\omega (q^2_\perp)=\frac{\kappa ^2}{(2\pi)^3}\,\int
\frac{q^2_\perp \,d^2k_\perp}{k^2_\perp (q-k)^2_\perp}\left(\frac{(k ,q-k )^2_\perp}{k_\perp^2}+
\frac{(k ,q-k )^2_\perp}{(q-k )^2_\perp}-q_\perp ^2-\frac{N}{2}\,(k ,q-k )_\perp\right).
\eeq
Here we added the contribution of $N$ gravitinos for the $N$-extended supergravity~\cite{Lip1}.
Other super-partners do not give any contribution in this order.

Note, that the infrared divergency of the Regge trajectory is universal, but the logarithmic
divergency at large $k_\perp$ depends on $N$ and is absent at $N=4$. Really the sum of the one-loop diagrams
do not contain any ultraviolet divergency, because the gravity is renormalized in one loop.
It means, that the integral over $k_\perp ^2$ is restricted from above by the value of
the order of $s$, which leads to the double-logarithmic asymptotics of the scattering amplitude
with the graviton quantum numbers in the $t$-channel. In Ref.~\cite{Lip1} the
corresponding ladder diagrams in the double-logarithmic approximation were summed and the following
result for the amplitude in the $N$-extended  supergravity was obtained
\beq
A_{2\rightarrow 2}=-\kappa ^2\,\frac{s^2}{t}\,\delta _{\lambda _A \lambda _{A'}}\delta _{\lambda _B \lambda _{B'}}
\frac{1}{a\xi}\,I_1(2a\xi)\,,
\eeq
where $\lambda_i$ are helicities of the initial and final gravitons, $I_n(x)$ is the modified Bessel function
and the parameters $a$ and $\xi$ are given below
\beq
a=\left((4-N)\,\frac{\kappa ^2}{16\pi^2}\,(-t)\right)^{\frac{1}{2}}\,,\,\,\xi =\frac{s}{t}\,.
\eeq
In principle there could be double-logarithmic contributions from other diagrams containing three and more
gravitons in the $t$-channel. To investigate this possibility one should generalize the effective action
constructed above to the supersymmetric case, because the contribution of the superpartners of the
graviton is essential for its Regge trajectory in higher loops. But we consider below for simplicity
only the first non-trivial correction to the action in the $N=1$ supergravity. In this case 
apart from the vierbein $e_\mu ^m$, related
to the metric tensor $g_{\mu \nu}$ in the well-known way
\beq
g_{\mu \nu}=\sum _n e_{\mu n}\,e_\nu ^n\,,
\eeq
the Rarita-Schwinger field $\psi _\mu$ describing the gravitino with the spin $3/2$ is introduced.
The action for this field is given below
\beq
S_{3/2}=\int d^4 x\,L_{3/2}\,\,,\,\,\,L_{3/2}=-\frac{1}{2}\epsilon ^{\mu \nu \rho \sigma}\,
\bar{\psi}_\mu \gamma_5\gamma_\nu D _\rho \psi _\sigma \,.
\eeq
The covariant derivative $D_\rho$ is defined by the relation
\beq
D_\rho =\partial_\rho +\frac{1}{2}\,\sigma _{m n}\,\omega ^{mn}_\rho \,,
\,\,\sigma _{m n}=\frac{1}{2}\,\left(\gamma _m\gamma_n-\gamma _n\gamma_m\right)\,,
\eeq
where $\omega ^{mn}_\rho$ is the spin connection expressed in terms of the Christoffel symbol
\[
\omega ^{mn}_\rho =-e^{\sigma n}\,\partial _\rho e^m_\sigma + e^{\sigma n}\,e^m_\alpha \,
\Gamma _{\rho \sigma}^\alpha
\]
\beq
=\frac{1}{2}\,e^{\sigma n}\,(\partial _\sigma e^m_\rho -\partial _\rho e^m_\sigma)-
\frac{1}{2}\,e^{\sigma m}\,(\partial _\sigma e^n_\rho -\partial _\rho e^n_\sigma)+
\frac{1}{2}\,e^{\nu n}\,e^{\mu m}\,e^{\rho k}\,(\partial _\nu e^k_\mu -\partial _\mu e^k_\nu)\,.
\eeq
The total action of supergravity is invariant under the supersymmetry transformation
\beq
\delta e_\mu ^m =\frac{\kappa}{2}\,\bar{\epsilon}\gamma ^m\psi_\mu \,,\,\,\delta \psi _\mu =\frac{1}{\kappa}\,
D_\mu \epsilon \,,
\eeq
where $\epsilon $ is a local parameter of these transformations being the anticommuting Majorano spinor.
It is known, that to close the SUSY commutator algebra off-shell one should introduce
the auxiliary fields $S,P$ and $A_m$. Here for simplicity of discussion we do not take into account them
neglecting total derivatives in the action and in its variation.

Let us  start again with the Born contribution to the induced contribution to the effective lagrangian
\beq
L_{ind}=-\frac{1}{4\kappa ^2}\,\left(j_{++}\partial_\sigma ^2A_{--}+j_{--}\partial_\sigma ^2A_{++}\right)\,,\,\,
j_{\pm \pm}\approx h_{\pm \pm}+...
\eeq
and attempt to add to it radiative corrections in the fields $h_{\mu \nu}$ and $\psi _\mu$ to derive its
generalization invariant under the
local supersymmetric transformations.

We obtain  the following infinitesimal transformation of the metric tensor with the light
cone components
\beq
\delta h_{\pm \pm}=\kappa \,\bar{\epsilon} \gamma_\pm \psi_\pm\,
\eeq
To cancel this term one should add to $j_{\pm \pm}$
 the contribution
 \beq
 \Delta _1 j_{\pm \pm}=\frac{\kappa^2}{2}\,\bar{\psi}_\pm \,\frac{\gamma_\pm}{\partial_\pm}\,\psi_\pm \,,
 \eeq
 because up to a total derivative in the integrand for the action its supersymmetric transformation
is equal to the expression
 \beq
 \delta_1 \left(\Delta _1 j_{\pm \pm}\right)\approx -\kappa \,\bar{\epsilon} \gamma_\pm \psi_\pm
 \eeq
opposite to $\delta h_{\pm \pm}$ in sign.

Thus, in the $N=1$ supersymmetric gravity we obtain for $j_{\pm \pm}$ with the next-to-leading accuracy
The following result
\beq
j_{\pm \pm}\approx h_{\pm \pm}-X_{\sigma \pm}^2+\frac{\kappa^2}{2}\,\bar{\psi}_\pm \,
\frac{\gamma_\pm}{\partial_\pm}\,\psi_\pm +...\,.
\eeq
The upper order corrections can be calculated in a similar way.

% The next-to-leading correction to the supersymmetric transformation of the
%gravitino contribution $\Delta _1 j_{\pm \pm}$ is
% \beq
%\delta_2 \left(\Delta _1 j_{\pm \pm}\right)=\kappa \,\frac{\omega _\pm ^{m n}}{4} \,
%\left(\bar{\psi}_\pm \frac{\gamma _\pm}{\partial _\pm}\,\sigma _{m n}\epsilon -\bar{\epsilon}\,
% \sigma _{m n}\,\frac{ \gamma_\pm}{\partial _\pm}\, \psi_\pm \right)\,.
% \eeq
% Also one obtains the following supersymmetric transformation of the next-to-leading term for $j_{\pm \pm}$
% \beq
% -\delta X_{\sigma \pm}^2=-X_{\sigma \pm}\,\kappa \,
% \left(\bar{\epsilon}\,\gamma _\sigma \psi_\pm +\bar{\epsilon}\,\gamma _\pm \psi_\sigma
% -\frac{\partial _\sigma}{\partial _\pm}\,\bar{\epsilon}\,\gamma _\pm \psi_\pm\right)
% \eeq

\section{Discussion}

In this paper the effective action for the high energy processes in gravity was constructed
in terms of the currents $j^\pm$ satisfying the Hamilton-Jacobi equation. This equation
can be solved in the perturbation theory or for simple configurations of the 
external gravitational fields. One can formulate a variational
principle for the currents calculated at such fields. The effective  action
can be used for the calculation of various elastic and inelastic scattering amplitudes in the Regge
kinematics. The Feynman rules for the simple vertices containing the reggeized gravitons are
extracted from the effective lagranjian. The one loop graviton Regge trajectory does not contain  
the ultraviolet divergency only in the N=4 supergravity. In other models the 
amplitudes with the graviton quantum numbers in the crossing channel have the double-logarithmic
terms. It is possible, that the constructed effective action can be generalized to the
case of superstrings living in the anti-de-Sitter 10-dimensional space. In this case one 
could use it for the discription of the Pomeron interactions at the N=4 supersymmetric gauge theory
in the framework of the AdS/CFT correspondence.

\section*{Acknowledgements}

I thank the Hamburg University for the hospitality
and J. Bartels, E. Levin, A. Sabio Vera and A. Prygarin for helpful discussions. This
work was supported by the grant RFFI-10-02-01338-a.

%\begin{verbatim}


\begin{thebibliography}{9}

\bibitem{Grib1} V.~N. Gribov, {\em Sov. Phys. JETP} {\bf 14}
478 (1962).

\bibitem{Mand} S. Mandelstam, {\em Nuovo Cim.} {\bf 30}, 1148 (1963).


\bibitem{Grib2} V.~N. Gribov, I.~Ya. Pomeranchuk and K.~A. Ter-Martirosyan,
{\em Phys.  Rev.} {\bf B 139}, 184 (1965).

\bibitem{Grib3} V.~N. Gribov, {\em Sov. Phys. JETP} {\bf 26},
414 (1968).

\bibitem{GGLMZ} M. Gell-Mann, M.~L. Goldberger, F.~E. Low, E. Marx and
F. Zachariasen, {\em Phys. Rev.} {\bf 133}, B145 (1954).

\bibitem{Mand1} S. Mandelstam, {\em Phys. Rev.} {\bf 137}, B949 (1965).

\bibitem{GST} M. T. Grisaru, H. J. Schnitzer and H.S. Tsao, {\em Phys. Rev. Lett.}
{\bf 30}, 811, (1973).

\bibitem{BFKL}
L.~N.~Lipatov,
{\em Sov.\ J.\ Nucl.\ Phys.}  {\bf 23} 338 (1976);\\
V.~S.~Fadin, E.~A.~Kuraev, L.~N.~Lipatov,
{\em Phys.\ Lett.} {\bf B 60} 50 (1975);\\
E.~A.~Kuraev, L.~N.~Lipatov, V.~S.~Fadin,
Sov.\ Phys.\ JETP {\bf 44} 443 (1976).


\bibitem{int1}
L.~N.~Lipatov,
{\em Phys.\ Lett.}  {\bf B 309} 394 (1993).

\bibitem{moeb}
L.~N.~Lipatov,
{\em Sov.\ Phys.\ JETP} {\bf 63} 904 (1986).

\bibitem{BKP}
Bartels, J.,
Nucl.\ Phys.\  {\bf B175} (1980) 365;\\
%%CITATION = NUPHA,B175,365;%%
Kwiecinskii, J., Praszalowicz, M.,
Phys.\ Lett.\  {\bf B94} (1980) 413.
%%CITATION = PHLTA,B94,413;%%

\bibitem{int}
L.~N.~Lipatov {\it
High energy asymptotics of
multi-colour QCD and exactly
solvable lattice models},
hep-th/9311037, unpublished.

\bibitem{dual}
L.~N.~Lipatov,
{\em Nucl.\ Phys.}  {\bf  B 548} 328 (1999).

\bibitem{FL}
V.~S.~Fadin, L.~N.~Lipatov,
{\em Phys.\ Lett.} {\bf B 429} 127 (1998);\\
M.~Ciafaloni and G.~Camici,
{\em Phys.\ Lett.}   {\bf B 430} 349 (1998).

\bibitem{trajN4}
A.~V.~Kotikov, L.~N.~Lipatov,
{\em Nucl.\ Phys.}  {\bf B 582} 19 (2000).

\bibitem{KL}
A.~V.~Kotikov, L.~N.~Lipatov,
{\em Nucl.\ Phys.} {\bf B 661} 19 (2003).


\bibitem{KLOV}
A.~V.~Kotikov, L.~N.~Lipatov, A.~I.~Onishchenko, V.~N.~Velizhanin,
{\em Phys.\ Lett.}  {\bf B 595} 521 (2004);
[Erratum-ibid.\  {\bf B 632} 754 (2006)].

\bibitem{integrQP}
L.~N.~Lipatov, talk at "Perspectives in Hadronic Physics", Proc. of Conf. ICTP,
Triest, Italy, May 1997.


\bibitem{Malda}
J.~M.~Maldacena,
{\em Adv.\ Theor.\ Math.\ Phys.}  {\bf 2} 231 (1998).


\bibitem{GKP}
S.~S.~Gubser, I.~R.~Klebanov, A.~M.~Polyakov,
{\em Phys.\ Lett.\  } {\bf B 428} 105 (1998).

\bibitem{W}
E.~Witten,
{\em Adv.\ Theor.\ Math.\ Phys.}  {\bf 2} 253 (1998).

\bibitem{BPST} R.~C.~Brower, J.~Polchinsky,  M.~J.~Strassler, C.~I.~Tan,
{JHEP} {\bf 0712} 005 (2007).

\bibitem{eff}
L.~N.~Lipatov,
{\em Nucl.\ Phys.}  B {\bf B 452}, 369 (1995);
{\em Phys.\ Rept.}  {\bf 286}, 131 (1997).

\bibitem{last}
E.~N.~Antonov, L.~N.~Lipatov, E.~A.~Kuraev, I.~O.~Cherednikov,
{\em Nucl.\ Phys.}  {\bf B 721}, 111 (2005).

\bibitem{BDS}
Z.~Bern, L.~J.~Dixon, V.~A.~Smirnov,
{\em Phys.\ Rev.}   {\bf D 72}, 085001 (2005).

\bibitem{BLS1}
J.~Bartels, L.~N.~Lipatov, A.~Sabio Vera, {\em Phys.\ Rev.} {\bf D 80}, 045002 (2009).

\bibitem{BLS2}
J.~Bartels, L.~N.~Lipatov, A.~Sabio Vera, {\em Eur.\ Phys.\ J.} {\bf C 65} 587, (2009),

\bibitem{Lip10}
L.~N.~Lipatov, preprint, hep-th 1008.1015.

\bibitem{LipPryg}
L.~N.~Lipatov, A.~Prygarin, preprints, hep-th 1008.1016, hep-th 1011.2673.

\bibitem{Intopen}
L.~N.~Lipatov,
{\em J. Phys.} {\bf A 42}, 304020 (2009).

\bibitem{GS} M.~T.~Grisaru, B.~van~Nieuwenhuizen and C.~C.~Wu, {\em Phys.\ Rev}
{\bf D 12} 1563; M.~T.~Grisaru, and H.~J.~Schnitzer, {\em Phys.\ Lett} {\bf 107B} 196 (1981).

\bibitem{Lip1} L.~N. Lipatov, {\em Phys. Lett.} {\bf 116B}, 411 (1982).

\bibitem{Lip2} L.~N. Lipatov, {\em JETP}, {\bf 82}, 991 (1982).

\bibitem{BAC} A.~Bellini, M.~Ademollo, M.~Ciafaloni, {\em Nucl.Phys.} {\bf B393}, 79
(1993).

\bibitem{Lip3} L.~N. Lipatov, {\em Nucl. Phys.} {\bf B365}, 614 (1991).

\bibitem{ACV} D.~Amati, M.~Ciafaloni and G.~Veneziano, {\em JHEP} {\bf 0802}, 049 (2008).

\bibitem{GSA} S.~B.~Giddings, M.~Schmidt-Sommerfeld and J.~R.~Andersen,  preprint,
hep-th 1005.5408.

\bibitem{Fock} V.~Fock, {\it The theory of space, time and gravitation}, Pergamon Press, London and Aylesbury
(1969).

\bibitem{Landau} L.~D.~Landau, E.~M.~Lifshitz, {\it Mechanics, Course of Theoretical Physics},
Butterworth Heinemann.

\bibitem{Schwarz} K.~Schwarzschild, S.~B.~Preuss. Acad. Wiss. 189 (1916).

\end{thebibliography}
\end{document}